\documentclass[aps,pre,twocolumn,superscriptaddress,nofootinbib]{revtex4}

\usepackage{graphicx}
\usepackage{amsmath}
\usepackage{amssymb}
\usepackage{color}
\usepackage{mathrsfs}
\usepackage[colorlinks=true,linkcolor=blue,citecolor=blue]{hyperref}
\usepackage[T1]{fontenc}

\begin{document}

\title{Time-fractional Caputo derivative versus other integro-differential
operators in generalized Fokker-Planck and generalized Langevin equations}

\author{Qing Wei}
\affiliation{School of Mechanics and Civil Engineering, China University of
Mining and Technology, Beijing, 10 0 083, PR China}
\affiliation{University of Potsdam, Institute of Physics \& Astronomy,
14476 Potsdam-Golm, Germany}
\author{Wei Wang}
\affiliation{University of Potsdam, Institute of Physics \& Astronomy,
14476 Potsdam-Golm, Germany}
\author{Hongwei Zhou}
\email{zhw@cumtb.edu.cn}
\affiliation{School of Energy and Mining Engineering, China University of
Mining and Technology, Beijing, 100083, PR China}
\author{Ralf Metzler}
\email{rmetzler@uni-potsdam.de}
\affiliation{University of Potsdam, Institute of Physics \& Astronomy,
14476 Potsdam-Golm, Germany}
\affiliation{Asia Pacific Center for Theoretical Physics, Pohang 37673,
Republic of Korea}
\author{Aleksei Chechkin}
\email{chechkin@uni-potsdam.de}
\affiliation{University of Potsdam, Institute of Physics \& Astronomy,
14476 Potsdam-Golm, Germany}
\affiliation{Faculty of Pure and Applied Mathematics, Hugo Steinhaus Center,
Wroclaw University of Science and Technology, Wyspianskiego 27, 50-370
Wroclaw, Poland}
\affiliation{Akhiezer Institute for Theoretical Physics National Science
Center, Kharkiv Institute of Physics and Technology, Akademichna 1, Kharkiv
61108, Ukraine}

\date{\today}

\begin{abstract}
Fractional diffusion and Fokker-Planck equations are widely used tools to
describe anomalous diffusion in a large variety of complex systems. The
equivalent formulations in terms of Caputo or Riemann-Liouville fractional
derivatives can be derived as continuum limits of continuous time
random walks and are associated with the Mittag-Leffler relaxation of
Fourier modes, interpolating between a short-time stretched exponential
and a long-time inverse power-law scaling. More recently, a number of
other integro-differential operators have been proposed, including the
Caputo-Fabrizio and Atangana-Baleanu forms. Moreover, the conformable
derivative has been introduced. We here study the dynamics of the associated
generalized Fokker-Planck equations from the perspective of the moments, the
time averaged mean squared displacements, and the autocovariance functions. We
also study generalized Langevin equations based on these generalized
operators. The differences between the Fokker-Planck and Langevin equations
with different integro-differential operators are discussed and compared
with the dynamic behavior of established models of scaled Brownian motion
and fractional Brownian motion. We demonstrate that the integro-differential
operators with exponential and Mittag-Leffler kernels are not suitable to
be introduced to Fokker-Planck and Langevin equations for the physically
relevant diffusion scenarios discussed in our paper. The conformable and
Caputo Langevin equations are unveiled to share similar properties with scaled
and fractional Brownian motion, respectively.
\end{abstract}

\maketitle

\section{Introduction}\label{section1}

Power-laws and fractional derivatives have a long tradition in the sciences.
Thus, Buelffinger modified Hooke's law to a general power exponent in 1729
\cite{buel,reiner}. Weber is credited for having (indirectly) discovered
viscoelasticity, following his report of a non-elastic aftereffect in stretched
silk threads in 1835 \cite{marko}, an effect that initiated the development of
generalized rheological models \cite{tschoegl}. An explicit time dependence
deviating from the exponential law was proposed by Kohlrausch in 1847 in the
form of a stretched exponential, introducing a fractional power-law into
an exponential function \cite{marko}. Power-law time dependencies in
relaxation phenomena were reported by Nutting in 1921 \cite{nutting}. To
grasp such a behavior in a compact dynamic equation, Scott Blair then
formulated a relaxation equation with a fractional derivative \cite{scott}.
Fractional rheological models have since been used widely to describe the 
viscoelastic or glassy behavior in complex systems \cite{wg,jcp95,helmut,
hilfer,mainardibook}. Fractional relaxation processes are also relevant in
biological contexts \cite{wg1,hilfer2000applications}.

Fractional derivatives have also found widespread use in the context of
anomalous diffusion, typically defined in terms of the power-law form
\cite{report,resto,pccp,hoefling,evangelista2018fractional}
\begin{equation}
\label{msd}
\langle x^2(t)\rangle\simeq K_{\alpha}t^{\alpha}
\end{equation}
of the mean squared displacement (MSD). Here $K_{\alpha}$ of physical
dimension $\mathrm{length}^2/\mathrm{time}^{\alpha}$ is the generalized
diffusion coefficient and the value of the anomalous diffusion exponent
$\alpha$ distinguishes subdiffusion ($0<\alpha<1$) from superdiffusion
($\alpha>1$), including the special cases of normal diffusion ($\alpha=1$)
and ballistic, wave-like motion ($\alpha=2$). A breakthrough in describing
anomalous diffusion came with the work of Schneider and Wyss \cite{schneider},
who started from the integral version of the diffusion equation
\begin{equation}
\label{intde}
P(x,t)-P_0(x)=K_1\int_0^t\frac{\partial^2}{\partial x^2}P(x,t')dt'
\end{equation}
for the probability density function $P(x,t)$ to find the test particle at
position $x$ at time $t$, given the initial condition $P_0(x)$ at $t=0$.
They then replaced the integral by a Riemann-Liouville (RL) fractional
integral operator defined for a suitable function $f(t)$ as \cite{oldham,
miller}
\begin{equation}
_0D_t^{-\alpha}f(t)=\frac{1}{\Gamma(\alpha)}\int_0^t\frac{f(t')}{(t-t')^{1
-\alpha}}dt',
\end{equation}
which is a direct generalization of the Cauchy multiple integral. After
differentiation by time once one obtains the equivalent fractional
diffusion equation (FDE) in RL-form \cite{report}
\begin{equation}
\label{fde}
\frac{\partial}{\partial t}P(x,t)=K_{\alpha}\,_0D_t^{1-\alpha}\frac{
\partial^2}{\partial x^2}P(x,t),
\end{equation}
where $_0D_t^{1-\alpha}=(d/dt)\,_0D_t^{-\alpha}$ is the RL fractional
differential operator \cite{oldham,miller}. The solution of the FDE (\ref{fde})
can be obtained in terms of Fox $H$-functions \cite{schneider,report,mathai}.
The asymptotic behavior of the PDF $P(x,t)$ encoded in the FDE (\ref{fde})
has a stretched Gaussian shape \cite{schneider,klazu,report}. FDEs of the
above form can be derived as the continuum limit of continuous time random
walks (CTRWs) with scale-free waiting time densities $\psi(t)\simeq t^{-1
-\alpha}$ with $0<\alpha<1$ and jump length densities with finite variance
\cite{compte,hilfer1995fractional,mebakla1,bakai2001,klazu,report,
thiel2014scaled}. By single-particle tracking method waiting time densities with
power-law forms were revealed, i.a., in protein motion in membranes \cite{weigel},
colloidal tracer motion in actin networks \cite{weitz,yael}, orfor tracers
in non-laminar flows \cite{swinney}. Power-law waiting time densities were also
identified in dynamic maps \cite{geisel} or simulations of drug molecules
diffusing in silica slits \cite{amanda}. When an external potential influences
the particle motion, the FDE (\ref{fde}) can be generalized to the
fractional Fokker-Planck equation \cite{mebakla,bakai2001,mebakla1,klazu,report,
magdziarz2007fractional,stanislavsky2008diffusion,gorska2012operator}. For
transport in groundwater and other applications advection terms as well as
mobile-immobile scenarios are considered in generalized versions of FDEs
\cite{schumer2009fractional,schumer2003fractal,timo,dentz2004time,
meerschaert2008tempered,harvey,brian,brian1,berkowitz2006modeling}. In the
context of FDEs and fractional Fokker-Planck equations of RL-type the
relaxation of modes follows the Mittag-Leffler pattern (see below) that
interpolates between an initial stretched exponential and a long-time
inverse power-law \cite{wg,report}.

Apart from the mentioned RL fractional derivatives, there exist a wide
variety of other types of fractional derivatives \cite{samko1993fractional,
podlubny1999fractional,kilbas2006theory,hilferdesi}, many of which are being
used in engineering and science applications. Possibly the most widely used
apart from the RL definition is the Caputo fractional derivative
\cite{caputo1966linear,caputo1967linear}. We note that both definitions are in
fact equivalent as long as the initial values are properly taken into account.
Thus, when we solve the FDE (\ref{fde}) for a specific initial value problem
the Caputo version of the FDE studied below leads to the same result. Caputo
fractional
derivatives are used, i.a., to model non-Darcian flow \cite{zhou2018fractional},
permeability models for rocks \cite{yang2019fractional}, contaminant transport
\cite{benson2013fractional}, or viscoelastic diffusion \cite{caputo2004diffusion,
caputo1999diffusion}. Apart from the Caputo or RL derivative based on a power-law
integral kernel with a (weak) singularity, in the last decade some new
non-singular integro-differential operators have been proposed. One option
is the Caputo-Fabrizio (CF) integro-differential operator with exponential
kernel \cite{caputo2015new}. The CF operator was employed in a number of areas,
for instance, fluid flow \cite{sheikh2018modern}, virus models
\cite{khan2019modeling, baleanu2020fractional}, and a human liver model
\cite{baleanu2020new}. Alternatively, the Atangana-Baleanu (AB)
integro-differential operator based on the Mittag-Leffler function for the
memory kernel \cite{atangana2016new} aims to describe the full memory effect
in systems since the Mittag-Leffler function combines a stretched exponential
shape and a power-law decay at short and long times, respectively. The AB
operator is used in FDEs \cite{sene2019analysis}, Cauchy and source problems
for advection-diffusion \cite{avci2019cauchy}, optimal control
\cite{ammi2019optimal}, and disease models \cite{baleanu2020modelling}.
Comparisons between the AB and CF integro-differential operators are
investigated in relaxation and diffusion models \cite{sun2017relaxation},
reaction-diffusion models \cite{shaikh2019analysis}, cancer models
\cite{gomez2017chaos}, heat transfer analysis \cite{siddique2021heat},
and for the Casson fluid \cite{ali2019effects}.

Apart from the non-local integro-differential operators mentioned above,
a local derivative, the so-called conformable derivative, has been introduced
\cite{A7,anderson2015newly,atangana2015new,fleitas2021note} and studied from
a physical point of view \cite{zhao2017general}. Applications of the
conformable derivative formulation have been discussed for anomalous
diffusion \cite{18}, advection-diffusion \cite{Y, 2020Analytical},
non-Darcian flow \cite{yang2018conformable}, and other differential equations
\cite{ccenesiz2017new,korpinar2019new,cevikelsolitary,akbulut2018auxiliary,
hyder2020exact}. A similar variant is the Hausdorff derivative proposed earlier
by Chen \cite{chen2006time}. This derivative is also employed in diffusion
scenarios \cite{chen2010anomalous,chen2017new,liang2019distributed,liang2019time,
liang2019hausdorff1}, anomalous diffusion in magnetic resonance imaging
\cite{liang2016fractal}, viscoelastic modeling \cite{cai2016characterizing},
or the Richards' equation \cite{sun2013fractal}. It was shown that the
conformable derivative is in fact proportional to the Hausdorff derivative
\cite{weberszpil2016variational,rosa2018dual}. Further discussions of these
derivatives can be found in \cite{diethelm2020fractional,giusti2018comment,
anderson2018nature}.

We here scrutinize the recently integro-differential operators proposed in
the framework of FDEs and generalized Fokker-Planck equations as well as their
applications in generalizations of the stochastic Langevin equations. Fourier
and Laplace transforms are used to obtain analytical solutions for the PDFs
and the moments to study the dynamics encoded in these dynamic equations. In
particular we unveil the connections between FDEs and fractional Langevin
equations with other well-known stochastic processes, particularly with scaled
Brownian motion (SBM, based on a Langevin equation with deterministic power-law
time dependence of the diffusion coefficient) and fractional Brownian motion
(FBM, based on a Langevin equation driven by zero-mean Gaussian noise with
long-range, power-law correlations). Our discussion is based
on experimentally measurable quantities. These include the first and second
moments, the MSD, and the PDF. Moments can be directly inferred from measured
time series as either ensemble or time averages \cite{barkai2012single}. PDFs
can also be reconstructed in many contemporary studies and used, inter alia,
to check for non-Gaussianity features \cite{ng}.

The paper is structured as follows. In Sec.~\ref{section2} we introduce and
briefly discuss different integro-differential
operators and recall the generalized Fokker-Planck and Langevin equations in
describing stochastic processes in complex systems. In Sec.~\ref{section3} we
discuss the results of local and non-singular integro-differential operators
in Fokker-Planck equations for the force-free case and for a constant drift.
Specifically we discuss the extent to which the CF and AB integro-differential
operators can provide physically meaningful descriptions in the anomalous
diffusion context. The results of the conformable and Caputo diffusion equations
(with drift) and SBM (with drift) are compared. In Sec.~\ref{section4} we focus
on generalized Langevin equations for SBM and FBM, as well as Langevin equations
with the four integro-differential operators introduced in Sec.~\ref{section2}.
Specifically, we again discuss the physical implications of the CF and AB
integro-differential operators in this context. The related moments, time
averaged MSD (TAMSD), and autocovariance function (ACVF) are considered to
assess different Langevin equations. In Sec.~\ref{section6}, we summarize and
discuss our results. We also present two tables with the main results for the
generalized Fokker-Planck and Langevin equations discussed in the paper.

\section{Integro-differential operators, generalized Fokker-Planck and Langevin equations}\label{section2}

In this Section, we provide a brief introduction to four definitions of
integro-differential operators that are frequently employed in theoretical
modeling and engineering, the Caputo derivative (note our remarks to the
extent these are equivalent to the RL-derivative) and conformable derivative,
as well as the two recently proposed CF and AB integro-differential
operators. We then introduce them to generalized formulations of the
Fokker-Planck and Langevin equations.

\subsection{Integro-differential operators}

The Caputo derivative of order $\alpha\in(0,1]$ for a suitable function
$f(t)$ is defined in terms of a power-law kernel \cite{caputo1966linear,
caputo1967linear},
\begin{equation}
\label{Caputodef}
^C_0D^{\alpha}_{t}f(t)=\frac{1}{\Gamma(1-\alpha)}\int_0^t\frac{f'(t')}{(t
-t')^{\alpha}}dt',
\end{equation}
where $f'(t)=df(t)/dt$.
The Caputo derivative thus has a (weak) singularity at $t'=t$. The special
feature introduced by Caputo in his derivative is the fact that the
derivative on the function $f(t)$, $f'(t)=df(t)/dt$, is contained
\emph{inside\/} the integral. This contrasts the definition of the RL
fractional derivative, in which the differentiation is taken \emph{after\/}
the fractional integration \cite{oldham,report}. In the Caputo formulation,
e.g., when using Laplace transform methods to derive the solution, the initial
conditions thus enter in the traditional way. For $0<\alpha<1$, e.g., the initial
value of $f(t)$ at $t=0$ is needed. This contrasts the RL derivative,
for which fractional-order initial conditions enter \cite{oldham}. However,
this complication for the RL derivative can be circumvented in the Schneider
and Wyss integral formulation presented above \cite{schneider,wg,wg1,report,resto,
mebakla,bakai2001,mebakla1}. In the solution of time-fractional equations with
initial condition given at $t=0$, the Laplace transform
\begin{equation}
\label{laplace}
\mathscr{L}\{f(t)\}(s)\equiv\tilde{f}(s)=\int_0^{\infty}f(t)\exp(-st)dt
\end{equation}
is of central importance. For the fractional RL-Integral, the Laplace transform
reads $\mathscr{L}\{f(t)\}=s^{-\alpha}\tilde{f}(s)$, and for the Caputo-fractional
operator we have
\begin{equation}
\mathscr{L}\left\{\,^C_0D^{\alpha}_{t}f(t)\right\}=s^{\alpha}\tilde{f}(s)-
s^{\alpha-1}f(0).
\end{equation}

Recently, two new definitions of integro-differential operators with
non-singular kernel were proposed. One is the Caputo-Fabrizio (CF)
integro-differential operator of order $\alpha\in(0,1]$ defined with
an exponential kernel \cite{caputo2015new},
\begin{equation}
\label{CFdef}
^{CF}_0D^{\alpha}_tf(t)=\frac{M(\alpha)}{(1-\alpha)\tau^\alpha}\int_0^t
f'(t')\exp\left(\frac{-\alpha(t-t')}{(1-\alpha)\tau}\right)dt',
\end{equation}
where the factor $M(\alpha)$ introduced in \cite{caputo2015new} is chosen
such that $M(0)=M(1)=1$. Note that we introduced the time scale $\tau$ in
order to get dimensions correct. The choice $\tau^{\alpha}$ in the factor
in front of the exponential allows us to take the limits $\alpha=0$ (where
we get $f(t)-f(0)$) and $\alpha=1$ (the normal differential) consistently.
The Laplace transform of the CF-operator reads
\begin{equation}
\mathscr{L}\left\{\,^{CF}_0D^{\alpha}_tf(t)\right\}=\frac{s\tilde{f}(s)-f(0)}{
(1-\alpha)\tau^{\alpha}s+\alpha\tau^{\alpha-1}}.
\end{equation}

The other variant for a generalized fractional derivative is given by the
Atangana-Baleanu (AB) integro-differential operator with Mittag-Leffler kernel
\cite{atangana2016new},
\begin{equation}
\label{ABdef}
^{AB}_0D^{\alpha}_tf(t)=\frac{B(\alpha)}{(1-\alpha)\tau^\alpha}\int_0^tf'(t')E_
\alpha\left(-\alpha\frac{(t-t')^\alpha}{(1-\alpha)\tau^\alpha}\right)dt',
\end{equation}
where $E_{\alpha}(-z)=\sum_{k=0}^{\infty}(-z)^k/\Gamma(1+\alpha k)$ is the
one-parameter Mittag-Leffler function with expansion around infinity $E_{\alpha}(-z)\sim-
\sum_{k=1}^{\infty}(-z)^{-k}/\Gamma(1-\alpha k)$ \cite{erdelyi}. In particular,
when $\alpha=1$,
$E_1(z)=e^z$. Note that we again introduced the time scale $\tau$ for
dimensional consistency. Moreover, we note that here $B(\alpha)$ is a normalization function satisfying $B(0)=B(1)=1$ and has the same properties as in the CF operator. To
simplify our notation in the following, we set $M(\alpha)=B(\alpha)=1$ \cite{caputo2015new, atangana2016new}.
The Laplace transform of the AB-operator has the form
\begin{equation}
\mathscr{L}\left\{\,^{AB}_0D^{\alpha}_tf(t)\right\}=\frac{s^{\alpha}\tilde{f}
(t)-s^{\alpha-1}f(0)}{(1-\alpha)\tau^{\alpha}s^{\alpha}+\alpha}.
\end{equation}

All these definitions, Caputo, CF, and AB integro-differential operators
correspond to convolutions of the derivative $f'(t)$ with different choices
for the kernels, i.e., power-law, exponential, and Mittag-Leffler functions,
respectively. In contrast, there also exists a local definition of a generalized
derivative, namely, the conformable derivative of order $\alpha\in(0,1]$,
defined via \cite{khalil2014new}
\begin{equation}\label{conformabledef}
T_{\alpha}f(t)=\lim\limits_{\bar{\tau}\to0}\frac{f(t+\bar{\tau}^{\alpha}t^{1
-\alpha})-f(t)}{\bar{\tau}^{\alpha}}.
\end{equation}
Here we used the small variable $\bar{\tau}$ with dimension of time to housekeep
physical dimensions. If the conformable derivative of $f$ of order
$\alpha$ exists in some interval $(0,a)$, $a>0$, and $\lim\limits_{t\to0^{+}}
T_{\alpha}f(t)$ exists, then we define $T_{\alpha}f(0)=\lim\limits_{t\to0^{+
}}T_{\alpha}f(t)$. The conformable Laplace transform of $f(t)$ is defined by
\cite{A7}
\begin{equation}
\label{CLT}
\mathscr{L}_{\alpha}\left\{f(t)\right\}(s)\equiv\tilde{f}_{\alpha}(s)=\int_0
^{\infty}f(t)t^{\alpha-1}\exp\left(-s\frac{t^{\alpha}}{\alpha}\right)dt,
\end{equation}
generalizing the standard Laplace transform (\ref{laplace}). The relationship
between the conformable Laplace transform and the ordinary Laplace transform is
$\mathscr{L}_{\alpha}\{f(t)\}(s)=\mathscr{L}\{f(\alpha t)^{1/\alpha)}\}(s)$.
The conformable Laplace transform of the conformable derivative is
\begin{equation}
\mathscr{L}_{\alpha}\{T_{\alpha}f(t)\}(s)=s\tilde{f}_{\alpha}(s)-f(0).
\end{equation}

The Hausdorff derivative (\emph{fractal\/} derivative) of a suitable function
$f(t)$ with respect to $t^\alpha$ \cite{chen2006time} is defined as
\begin{equation}
\label{fractalderivative}
\frac{df(t)}{dt^\alpha}=\lim_{t'\rightarrow t}\frac{f(t)-f(t')}{t^\alpha
-(t')^\alpha}.
\end{equation}
From this definition we see that the Hausdorff derivative is also local in
nature. The connection between the Hausdorff derivative and the conformable
derivative is given by \cite{weberszpil2016variational,rosa2018dual},
\begin{equation}
\label{Con-fractal}
\alpha\frac{df(t)}{dt^\alpha}=t^{1-\alpha}\frac{df(t)}{dt}=T_{\alpha}f(t).
\end{equation}

\subsection{Generalized Fokker-Planck equations}

We now use these operators to generalize the Fokker-Planck equation
\cite{risken}
\begin{eqnarray}
\label{FPE}
\frac{\partial}{\partial t}P(x,t)=\left(K_1\frac{\partial^2}{\partial x^2}
-\frac{\partial}{\partial x}\frac{F(x)}{m\eta}\right)P(x,t)
\end{eqnarray}
in the presence of a general external force field $F(x)$. Here $m$ is the
particle mass and $\eta$ the friction coefficient \cite{risken}. For
the force, we obtain solutions for vanishing or constant external force
$F_0$. In the latter case, we then use the drift velocity $v=F_0/(m\eta)$.
In the generalization based on the RL derivative the fractional Fokker-Planck
equation was analyzed for different linear and non-linear force fields
\cite{bakai2001,mebakla,report,resto}.

We here introduce the four generalized differential operators above and
analyze the generalized diffusion equation (GDE) or generalized diffusion
equation with drift (drift-GDE)
\begin{equation}
\label{ffpe}
\frac{\partial^{\alpha}}{\partial t^{\alpha}}P(x,t)=\left(K_{\alpha}\frac{
\partial^2}{\partial x^2}-v_{\alpha}\frac{\partial}{\partial x}\right)P(x,t)
\end{equation}
with initial condition $P(x,0)=\delta(x)$, for $v_{\alpha}=0$ and $v_{\alpha}
\neq0$ and "natural" boundary conditions $P(|x|\to\infty,t)=0$. These standard
initial and boundary conditions will be applied throughout this work. Note that
we introduced the $\alpha$-dependent velocity $v_{\alpha}$
for dimensionality housekeeping purposes. This can be achieved by setting
$v_{\alpha}=v\tau^{1-\alpha}$, where $\tau$ is a time scale, such that $v_{
\alpha}$ has dimension $\mathrm{length}/\mathrm{time}^{\alpha}$. We seek
solutions of Eq.~(\ref{ffpe}) on the infinite line $-\infty<x<\infty$, and the
notation $\partial ^{\alpha}/\partial t^{\alpha}$ represents our four operators.

For our analysis, we also need to introduce SBM \cite{lim2002self,
jeon2014scaled}, whose diffusion coefficient is explicitly time-dependent
and evolves as power-law $\mathscr{K}_\alpha(t)=\alpha K_\alpha t^{\alpha-1}$
with $\alpha>0$. SBM is a Gaussian self-similar Markovian process with independent but non-stationary increments. It finds applications in turbulence \cite{batchelor1952}, stochastic hydrology \cite{talkner2011}, finance \cite{bassler2007}, granular gases \cite{bodrova2015}, and magnetic resonance imaging \cite{novikov2014}, to name a few. The Fokker-Planck equation for SBM in an external force
field is given by \cite{jeon2014scaled}
\begin{eqnarray}
\label{FPESBM}
\frac{\partial}{\partial t}P(x,t)=\left(\mathscr{K}_\alpha(t)\frac{\partial^2}
{\partial x^2}-\frac{\partial}{\partial x}\frac{F(x)}{m\eta}\right)P(x,t).
\end{eqnarray}

\subsection{Generalized Langevin equations}

Fokker-Planck equations are deterministic equations for the PDF $P(x,t)$. A
stochastic description of the position of a test particle in the presence of a
fluctuating force is the Langevin equation \cite{coffey}, the alternative 
standard description of diffusive processes \cite{vankampen}. The (overdamped)
Langevin equation corresponding to the Fokker-Planck equation (\ref{FPE})
with $P_0(x)=\delta(x)$ reads
\begin{equation}
\label{lan}
\frac{d}{dt}x(t)=\frac{F(x)}{m\eta}+\sqrt{2K_1}\xi(t),
\end{equation}
with the initial position $x(0)=0$. Here $\xi(t)$ is zero-mean white Gaussian
noise with ACVF $\langle\xi(t)\xi(t')\rangle=\delta(t-t')$.
Then, the \textit{generalized} Langevin equation for the thermalized system in the overdampeld approximation reads \cite{zwanzig2001}
\begin{equation}\label{kernelLE}
\int_0^t \gamma\left(t-t^{\prime}\right) \frac{\mathrm{d} x\left(t^{\prime}\right)}{\mathrm{d} t^{\prime}} \mathrm{d} t^{\prime}=\frac{F(x)}{m\eta}+\frac{\zeta(t)}{m\eta}, 
\end{equation}
where the noise autocorrelation function is coupled to the frictional kernel by the Kubo-Zwanzig fluctuation-dissipation theorem (FDT) \cite{kubo1966} $\left\langle\zeta(t_1)\zeta(t_2)\right\rangle=k_{B}Tm\eta\gamma\left(t_1-t_2\right)$, where $k_B$ is the Boltzmann constant and $T$ is the absolute temperature of the environment. The noise $\zeta(t)$ obeying the FDT is called \textit {internal}. Generalized Langevin equations with FDT and different kernels, e.g., of exponential and Mittag-Leffler shapes have been extensively studied before \cite{vinales2007, figueiredo2009,liemert2017,fa2006, porra1996}.

In what follows we
generalize the Langevin equation (\ref{lan}) via the four
operators in Eqs. (\ref{Caputodef}), (\ref{CFdef}), (\ref{ABdef}), and (\ref{conformabledef}), in presence of a constant force, with the unifying notation
\begin{equation}
\label{langevin-eqaution}
\frac{d^{\alpha}}{d t^{\alpha}}x(t)=v_{\alpha}+\sqrt{2K_{\alpha}}\xi(t),
\end{equation}

We note that FBM and the generalized Langevin equations we consider here do not
fulfill the generalized fluctuation-dissipation theorem and thus do not describe
equilibrium systems. Instead, the noise is considered to be \emph{external}
\cite{klimo}. This is appropriate for active systems, in which energy is
dissipated, e.g., living biological cells.

We will compare the resulting dynamics with that encoded by SBM, whose
Langevin equation is given by \cite{lim2002self,jeon2014scaled}
\begin{eqnarray}
\label{LESBM}
\frac{d}{dt}x(t)=\frac{F(x)}{m\eta}+\sqrt{2\mathscr{K}_\alpha(t)}\xi(t),
\end{eqnarray}
which is equivalent to the deterministic equation (\ref{FPESBM}). The noise
$\xi(t)$ has the same properties as for the standard Langevin equation
(\ref{lan}), i.e., it is zero-mean white Gaussian noise. In the following we
will consider the cases of zero and constant force.

As we will see, it will also be of interest to compare our results to the
dynamics of FBM \cite{mandelbrot1968fractional}, which is stationary in
increments and nearly ergodic \cite{deng2009ergodic,pre12,lene1}. As a
generalization of Brownian motion, FBM is an effective stochastic process
to model anomalous diffusion \cite{barkai2012single}. Its Langevin equation
reads \cite{mandelbrot1968fractional,pccp}
\begin{eqnarray}
\label{LEFBM}
\frac{d}{dt}x(t)=\frac{F(x)}{m\eta}+\sqrt{2K_{\alpha}}\xi_\alpha(t),
\end{eqnarray}
where $\xi_\alpha(t)$ is zero-mean fractional Gaussian noise with the long
range, power-law ACVF $\langle\xi_\alpha(t_1)\xi_\alpha(t_2)\rangle\sim
\alpha(\alpha-1)|t_1-t_2|^{\alpha-2}$ when $|t_1-t_2|\gg1$. FBM is defined
for the range $0<\alpha<2$ of the anomalous diffusion exponent, instead of
which the Hurst exponent $H=\alpha/2$ is often used. From the noise ACVF we
can see that the noise correlations are positive (persistent) when the
motion is superdiffusive, while they are negative (antipersistent) in the
subdiffusive case.

We will characterize the dynamics of the processes we consider the
moments, the time-averaged MSD (TAMSD) \cite{barkai2012single,pccp}, and the
displacement ACVF of the process. The TAMSD is important in the analysis of
single particle trajectories measured in modern tracking experiments; it is
defined via
\begin{equation}
\label{TAMSD}
\overline{\delta^2(\Delta)}=\frac{1}{T-\Delta}\int_0^{T-\Delta}\Big(x(t+\Delta)
-x(t)\Big)^2dt,
\end{equation}
where $T$ is the length of the time series (measurement time) and $\Delta$ is
called the lag time. The mean TAMSD is obtained from averaging over a number
$N$ of individual traces $\overline{\delta_i^2(\Delta)}$,
\begin{equation}
\left<\overline{\delta^2(\Delta)}\right>=\frac{1}{N}\sum_{i=1}^N
\overline{\delta_i^2(\Delta)}.
\end{equation}
In the Birkhoff-Boltzmann sense, a system is considered ergodic when ensemble
and time averages are equivalent in the limit of long measurement times. We
here consider a stochastic process non-ergodic when the ensemble-averaged
MSD and the TAMSD are disparate in the limit
of long observation times, $\lim\limits_{T\rightarrow\infty} \overline{
\delta^2(\Delta)}\neq\langle x^2(\Delta)\rangle$. The displacement ACVF is
defined as
\begin{equation}
\label{acvfdef}
C_{\Delta}(t)=C_{\Delta}(t)=\frac{\langle[x(t+\Delta)-x(t)][x(\Delta )
-x(0)]\rangle}{\Delta^2}.
\end{equation}

\section{Generalized Fokker-Planck equations}
\label{section3}

\subsection{Generalized Fokker-Planck equations in the force-free case}
\label{section3A}

We now first assess the dynamics encoded in the generalized Fokker-Planck
equation (\ref{ffpe}) in the absence of an external force.

\subsubsection{Caputo and non-singular differential operators}
\label{subsection3B}

When $F(x)=0$, the generalized Fokker-Planck equation (\ref{ffpe}) reduces
to the GDE
\begin{equation}
\label{eqd}
\frac{\partial^{\alpha}}{\partial t^{\alpha}}P(x,t)=K_{\alpha}\frac{\partial^2}
{\partial x^2}P(x,t),
\end{equation}
with $0<\alpha\le1$ which we will solve on the interval $-\infty<x<\infty$ for
the initial condition $P(x,0)=\delta(x)$. Here the operator $\partial^{\alpha}/
\partial t^{\alpha}$ represents the Caputo, CF, and AB integro-differential
operators, Eqs.~(\ref{Caputodef}), (\ref{CFdef}), and (\ref{ABdef}).

Applying the Laplace and Fourier transforms to the GDE (\ref{eqd}), we find
\begin{eqnarray}
\label{C1}
\hat{\tilde{P}}_C(k,s)&=&\frac{s^{\alpha-1}}{s^{\alpha}+K_{\alpha}k^2},\\
\label{eq3}
\hat{\tilde{P}}_{CF}(k,s)&=&\frac{1}{s+K_{\alpha}[\alpha\tau^{\alpha-1}+s(1-
\alpha)\tau^\alpha]k^2},\\
\label{eq4}
\hat{\tilde{P}}_{AB}(k,s)&=&\frac{s^{\alpha-1}}{s^{\alpha}+K_{\alpha}[\alpha
+s^{\alpha}(1-\alpha)\tau^{\alpha}]k^2}.
\end{eqnarray}
After inverse Fourier transform we obtain the PDFs in Laplace space,
\begin{equation}
\label{lapsol}
\tilde{P}(x,s)=\frac{1}{2s}\phi(s)\exp\left(-\phi(s)|x|\right),
\end{equation}
where we define
\begin{eqnarray}
\label{eqCaputo}
\phi_C(s)&=&\sqrt{\frac{s^\alpha}{K_{\alpha}}},\\
\label{eqs}
\phi_{CF}(s)&=&\sqrt{\frac{s}{(1-\alpha)\tau^{\alpha}K_{\alpha}s+\alpha\tau^{
\alpha-1}K_{\alpha}}},\\
\label{eqs1}
\phi_{AB}(s)&=&\sqrt{\frac{s^{\alpha}}{(1-\alpha)\tau^{\alpha}K_{\alpha}
s^{\alpha}+\alpha K_{\alpha}}},
\end{eqnarray}
for the three operators, respectively.

We first recall the PDF of the Caputo diffusion equation based on the Caputo
operator (\ref{Caputodef}) in $(x,t)$-space,
\begin{equation}
\label{C2}
P_C(x,t)=\frac{1}{2\sqrt{K_{\alpha}t^{\alpha}}}M_{\alpha/2}\left(\frac{|x|}{
\sqrt{K_{\alpha} t^{\alpha}}}\right),
\end{equation}
where $M_\nu$ denotes the Mainardi function \cite{mainardi2003wright,
gorenflo2007analytical}, also called $M$ function (of the Wright type), of
order $\nu$. Using the asymptotic representation of $M_{\alpha/2}$ the PDF
$P_C(x,t)$ can be shown to have a stretched Gaussian shape
\cite{mainardi2007fundamental}, corresponding to the above-mentioned results
in \cite{schneider,klazu,report}. An alternative representation of the PDF
is via Fox $H$-functions \cite{schneider,report}. The MSD of the Caputo-FDE
reads
\begin{equation}
\label{eq11}
\langle x^2(t)\rangle_C=\frac{2K_{\alpha}}{\Gamma(\alpha+1)}t^{\alpha},
\end{equation}
and the corresponding kurtosis has the time-independent value
\begin{equation}
\label{KC}
\kappa_C=\frac{3\alpha\Gamma(\alpha)^{2}}{\Gamma(2\alpha)},
\end{equation}
demonstrating the non-Gaussian character of $P_C(x,t)$.

To vouchsafe that the solutions of the AB-DE and the CF-DE are proper PDFs,
they should be completely monotonic in the Laplace domain \cite{Rschilling2010}.
That is, we should first check the complete monotonicity of the form
(\ref{lapsol}) together with Eqs.~(\ref{eqs})
and (\ref{eqs1}). In App.~\ref{AppendixB1}, we use the theory of Bernstein
functions to demonstrate that the complete monotonicity is indeed ensured.

We now calculate the long-time limit of the PDF for the GDEs of CF and AB
type from the Laplace representation in Eq.~(\ref{lapsol}) together with
Eqs.~(\ref{eqs}), and (\ref{eqs1}). In Laplace space, the long-time limit
$t\to\infty$ corresponds to $s\to0$, for which we find the asymptotic
behaviors
\begin{eqnarray}
\nonumber
\tilde{P}_{CF}(x,s)&\sim&\frac{1}{2s}\sqrt{\frac{s}{\alpha\tau^{\alpha-1}
K_{\alpha}}}\\
&&\times\exp\left(-\sqrt{\frac{s}{\alpha\tau^{\alpha-1}K_{\alpha}}}|x|\right),\\
\tilde{P}_{AB}(x,s)&\sim&\frac{1}{2s}\sqrt{\frac{s^\alpha}{\alpha K_{\alpha}}}
\exp\left(-\sqrt{\frac{s^\alpha}{\alpha K_{\alpha}}}|x|\right).
\end{eqnarray}
Laplace back-transforming, these forms correspond to the long-time asymptotes
\begin{eqnarray}
\nonumber
P_{CF}(x,t)&\sim&\frac{1}{\sqrt{4\alpha\tau^{\alpha-1} K_\alpha\pi t}}\\
\label{CFDElong}
&&\times\exp\left(-\frac{x^2}{4\alpha\tau^{\alpha-1}K_{\alpha}t}\right),\\
\label{ABDElong}
P_{AB}(x,t)&\sim&\frac{1}{\sqrt{4\alpha K_\alpha t^\alpha}} M_{\alpha/2}
\left(\frac{|x|}{\sqrt{\alpha K_\alpha t^\alpha}}\right).
\end{eqnarray}
Eq.~(\ref{CFDElong}) demonstrates that the CF-DE describes normal diffusion
at long times, with a Gaussian shape and $x$ scaling like $t^{1/2}$. In contrast
Eq.~(\ref{ABDElong}) for the AB-DE at long times is similar to the PDF of the
Caputo-FDE. Both behaviors are expected from the definitions of the respective
integro-differential operators: at long times, the CF-operator includes an
exponential cutoff, whereas the AB-operator has an asymptotic power-law tail
equivalent to the kernel in the Caputo operator.

The short-time limit of the PDFs for the CF- and AB-DEs can be similarly
calculated from their Laplace representations (\ref{lapsol}), (\ref{eqs}),
and (\ref{eqs1}).
For $t\to0$, corresponding to $s\to\infty$ in the Laplace domain, we have
\begin{eqnarray}
\nonumber
\tilde{P}_{CF}(x,s),\tilde{P}_{AB}(x,s)&\sim&\frac{1}{2s\sqrt{(1-\alpha)
\tau^{\alpha}K_{\alpha}}}\\
&&\hspace*{-2.4cm}\times\exp\left(-\frac{|x|}{\sqrt{(1-\alpha)\tau^{\alpha}K_{
\alpha}}}\right),
\end{eqnarray}
from which we obtain the asymptotic short-time result in the time domain,
\begin{eqnarray}
\nonumber
P_{CF}(x,t),P_{AB}(x,t)&\sim&\frac{1}{\sqrt{4(1-\alpha)\tau^{\alpha}K_{\alpha}}}\\
&&\hspace*{-2.4cm}\times\exp\left(-\frac{|x|}{\sqrt{(1-\alpha)\tau^{\alpha}
K_{\alpha}}}\right),
\label{funpdf}
\end{eqnarray}
This is a remarkable result, showing a finite width of the initial condition,
which we will comment on below. The PDFs of the Caputo-, CF-, and AB-DEs are
shown for different times in Fig.~\ref{Fig1}a-c. Indeed, for the CF and AB
cases the shapes of the limits for $t=0$ correspond to a Laplace distribution.

We now calculate the MSDs for the CF- and AB-DEs,
\begin{equation}
\label{eq12}
\langle x^2(t)\rangle_{CF}=2\alpha K_{\alpha}\tau^{\alpha-1}t+2(1-\alpha)
K_{\alpha}\tau^{\alpha},
\end{equation}
and
\begin{equation}
\label{eq13}
\langle x^2(t)\rangle_{AB}=\frac{2K_{\alpha}}{\Gamma(\alpha)}t^{\alpha}+2(1
-\alpha)K_{\alpha}\tau^{\alpha}.
\end{equation}
While the long time behaviors $\langle x^2(t)\rangle_{CF}\simeq t$ and $\langle
x^2(t)\rangle_{AB}\simeq
t^{\alpha}$ produce normal and subdiffusive scaling, the values of the MSDs in
the limit $t\to0$ have finite values, corresponding to the finite width of the
limits (\ref{funpdf}) of the PDFs. We also calculated the kurtosis in Appendix
\ref{AppendixB1}, Eqs.~(\ref{KCF}) and (\ref{KAB}). At short times, $\kappa_{CF},
\kappa_{AB}\sim6$, this means the CF- and AB-DE describe non-Gaussian process. At
long times, $\kappa_{CF}\sim3$, i.e., the CF-DE describes a Gaussian process,
while $\kappa_{AB}\sim3\alpha\Gamma(\alpha)^2/\Gamma(2\alpha)=\kappa_C$, which
shows that the AB-DE is similar to the PDF of the Caputo-FDE in this long time
limit. The kurtosis of these models are shown in Fig.~\ref{Fig2}. The MSDs
of the GDE with Caputo-, CF-, and AB-operators are shown in Fig.~\ref{Fig20}.

From this discussion we see that within the framework of the GDE considered
here with initial value $P(x,0)$ given at time $t=0$ the CF- and AB-operators
produce inconsistent results in the short time limit. This observation
deserves a separate formal investigation. We note that the same results
for the Fourier-Laplace forms Eqs.~(\ref{eq3}) and (\ref{eq4}) can be derived
from the corresponding integral formulations (see Appendix \ref{AppendixD}) of these operators.

Next we establish the relation of CF-DE and AB-DE in Eq.~(\ref{eqd}), to the continuous time random walk.

\subsubsection{Relation to continuous time random walk}

The generalized diffusion equation arises as a long space-time limit of CTRW, characterized by two PDFs, the distributions of jumps $\lambda(x)$ and waiting times $\psi(t)$. The jump distribution possesses a finite variance, and this property leads to the appearance of the second order space derivative on the right-hand side of the generalized diffusion equation. The waiting time distributions determines the kernel $\theta(t)$ of the integro-differential operator on the left-hand side. In Laplace space, this relation obtains a simple form, see, e.g. \cite{sandev2015},
\begin{equation}\label{47}
\widetilde{\psi}(s)=\frac{1}{1+\tau^\alpha s \tilde{\theta}(s)},
\end{equation}
The waiting time PDF together with a Gaussian jump length PDF with $\hat{\lambda}(k) \sim 1-\sigma^2 k^2$ yield the Fourier-Laplace form of the Montroll-Weiss relation
\begin{equation}\label{pks}
    \hat{\tilde{P}}(k, s)=\frac{1-\widetilde{\psi}(s)}{s}\frac{1}{1-\widetilde{\psi}(s)(1-\sigma^2 k^2)},
\end{equation}
for the PDF. 
Rewriting Eq.~(\ref{pks}) as 
Eq.~(\ref{rpks})
\begin{equation}\label{rpks}
    \tilde{\theta}(s)[s \hat{\tilde{P}}(k, s)-1]=-\frac{\sigma^2}{\tau^{\alpha}} k^2 \hat{\tilde{P}}(k, s),
\end{equation}
and taking on inverse Fourier-Laplace transform we obtain the generalized diffusion equation
\begin{equation}
    \int_0^t \theta\left(t-t^{\prime}\right) \frac{\partial}{\partial t^{\prime}} P\left(x, t^{\prime}\right) \mathrm{d} t^{\prime}=K_{\alpha}\frac{\partial^2}{\partial x^2} P(x, t),
\end{equation}
with the memory kernel $\theta(t)$, and $K_{\alpha}=\sigma^2/\tau^{\alpha}$. The initial condition is again of the form $P_0(x)=\delta(x)$, i.e. $\hat{P}_0(k)=1$.

For the Caputo derivative, CF and AB operators, see Eqs.~(\ref{Caputodef}), (\ref{CFdef}) and (\ref{ABdef}), the corresponding kernels in the GDEs have the form 
\begin{eqnarray}
\label{kernel}
\nonumber
\theta_{C}(t)&=&\frac{t^{-\alpha}}{\Gamma(1-\alpha)},\\
\nonumber
\theta_{CF}(t)&=&\frac{1}{(1-\alpha)\tau^{\alpha}}\exp{\left(-\frac{\alpha t}{(1-\alpha)\tau}\right)}, \\
\theta_{AB}(t)&=&\frac{1}{(1-\alpha)\tau^{\alpha}}E_{\alpha}\left(-\alpha\frac{t^{\alpha}}{(1-\alpha)\tau^{\alpha}}\right).
\end{eqnarray}

We here recall the waiting time PDF for the Caputo FDE. In Laplace space, 
\begin{equation}
\widetilde{\psi}_{C}(s)=\frac{1}{\tau^\alpha s^{\alpha}+1},
\end{equation}
which is a completely monotonic function \cite{sandev2015}.
Then the corresponding waiting time PDF is 
\begin{equation}
\psi_{C}(t)=\frac{1}{\tau^{\alpha}}t^{\alpha-1}E_{\alpha,\alpha}\left(-\frac{t^{\alpha}}{\tau^{\alpha}}\right),
\end{equation}
where $E_{\alpha, \beta}(-z)=\sum_{k=0}^{\infty}(-z)^k/\Gamma(\beta+\alpha k)$ is the
two-parameter Mittag-Leffler function with expansion around infinity $E_{\alpha, \beta}(-z)\sim-\sum_{k=1}^{\infty}(-z)^{-k}/\Gamma(\beta-\alpha k)$ \cite{erdelyi}. In particular,
when $\beta=1$,
$E_{\alpha,1}(z)=E_{\alpha}(z)$. We note that $\psi_{C}(t)$ has a weak singularity at $t=0$, and in the long time limit, with $-z\Gamma(-z)\Gamma(z)=\pi\csc(\pi z)$ \cite{abramowitz1964handbook}, we have 
\begin{eqnarray}
\nonumber
\lim_{t\to+\infty}\psi_{C}(t)&&=\lim_{t\to+\infty}\frac{t^{\alpha-1}}{\tau^{\alpha}}E_{\alpha,\alpha}\left(-\frac{t^{\alpha}}{\tau^{\alpha}}\right)\\
&&\sim \frac{\Gamma(\alpha+1)\sin(\pi\alpha)\tau^{\alpha}}{\pi }t^{-(1+\alpha)},
\end{eqnarray}
that is, at long times, $\psi_{C}(t)$ decays as $t^{-(1+\alpha)}$.
We note that $\psi_{C}(t)$ satisfies normalization, i.e.,
\begin{eqnarray}
\nonumber
\int_0^{\infty}\psi_{C}(t')dt'&&=\frac{1}{\tau^{\alpha}}\int_0^{\infty}t'^{\alpha-1}E_{\alpha,\alpha}\left(-\frac{t'^{\alpha}}{\tau^{\alpha}}\right)dt'\\
\nonumber
&&=\lim_{z\to+\infty} \frac{1}{\tau^{\alpha}}z^{\alpha}E_{\alpha,\alpha+1}\left(-\frac{z^{\alpha}}{\tau^{\alpha}}\right)\\
&&=1.
\end{eqnarray}

Now we consider the waiting time PDF for CF and AB cases, first in Laplace space. According to Eqs.~(\ref{47}) and (\ref{kernel}), we obtain 
\begin{equation}
\widetilde{\psi}_{CF}(s)=\frac{1-\alpha}{2-\alpha}+\frac{\alpha}{2-\alpha}\frac{1}{(2-\alpha)\tau s+\alpha},
\end{equation}
\begin{equation}
\widetilde{\psi}_{AB}(s)=\frac{1-\alpha}{2-\alpha}+\frac{\alpha}{2-\alpha}\frac{1}{(2-\alpha)\tau^{\alpha} s^{\alpha}+\alpha}.
\end{equation}
Notice that $\widetilde{\psi}_{CF}(s)$ and $\widetilde{\psi}_{AB}(s)$ are completely monotonic functions, therefore by applying an inverse Laplace transform we obtain functions, which are proper PDFs in time domain, 
\begin{equation}\label{CFCTRWkernel}
\psi_{CF}(t)=\frac{1-\alpha}{2-\alpha}\delta(t)+\frac{\alpha}{\tau(2-\alpha)^{2}}\exp\left(-\frac{\alpha}{\tau(2-\alpha)}t\right),
\end{equation}
\begin{equation}\label{ABCTRWkernel}
\psi_{AB}(t)=\frac{1-\alpha}{2-\alpha}\delta(t)+\frac{\alpha t^{\alpha-1}}{\tau^{\alpha}(2-\alpha)^{2}}E_{\alpha,\alpha}\left(-\frac{\alpha}{\tau^{\alpha}(2-\alpha)}t^{\alpha}\right).
\end{equation}
One can easily check their normalization, $\int_0^{\infty}\psi_{CF}(t')dt'=1$, $\int_0^{\infty}\psi_{AB}(t')dt'=1$.

From Eqs.~(\ref{CFCTRWkernel}) and (\ref{ABCTRWkernel}), we notice that the waiting time PDFs of both two cases contain the term $\delta(t)$, which means that the particle jumps at the initial time $t=0$, instead of waiting on site. This observation is in line with the property of having a non-zero MSD at $t=0$, see Eqs.~(\ref{eq12}) and (\ref{eq13}).
It may imply that the use of CF and AB operators in anomalous dynamics requires proper initial conditions
that are different from those used in standard formulations of the diffusion problem (i.e., $P(x,t=0)=\delta(x)$)
and in standard formulations of the continuous time random walk models, in which the particle arrives at a site at $t=0$, then waits and then makes a jump. Such a discussion goes beyond the scope of our paper and requires further investigation. Here we only conclude that while CF and AB integro-differential operators may be useful for the description of other generalized dynamics, we do not consider them for the  following discussion of GDEs.
Below we show that these two operators also lead to inconsistent
formulations of the generalized Langevin equations.

\subsubsection{Alternative formulation for the CF-GDE}

\begin{figure*}
(a)\includegraphics[width=7.2cm]{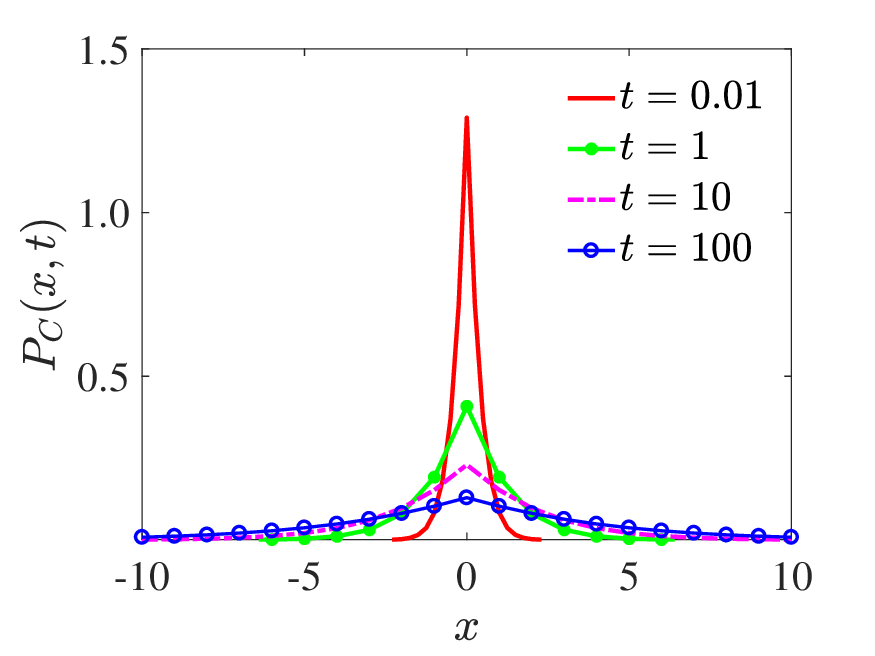}
\includegraphics[width=7.2cm]{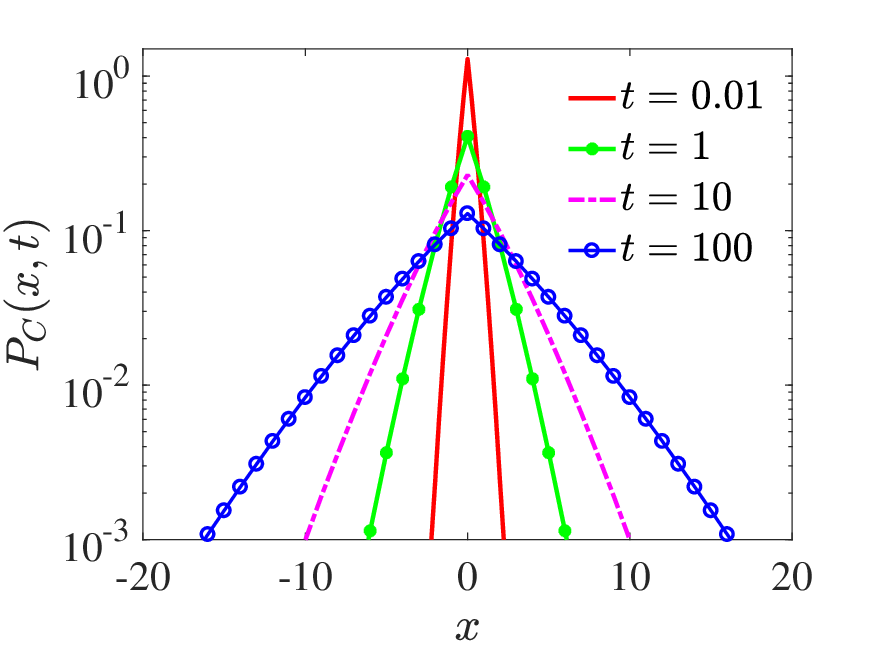}
(b)\includegraphics[width=7.2cm]{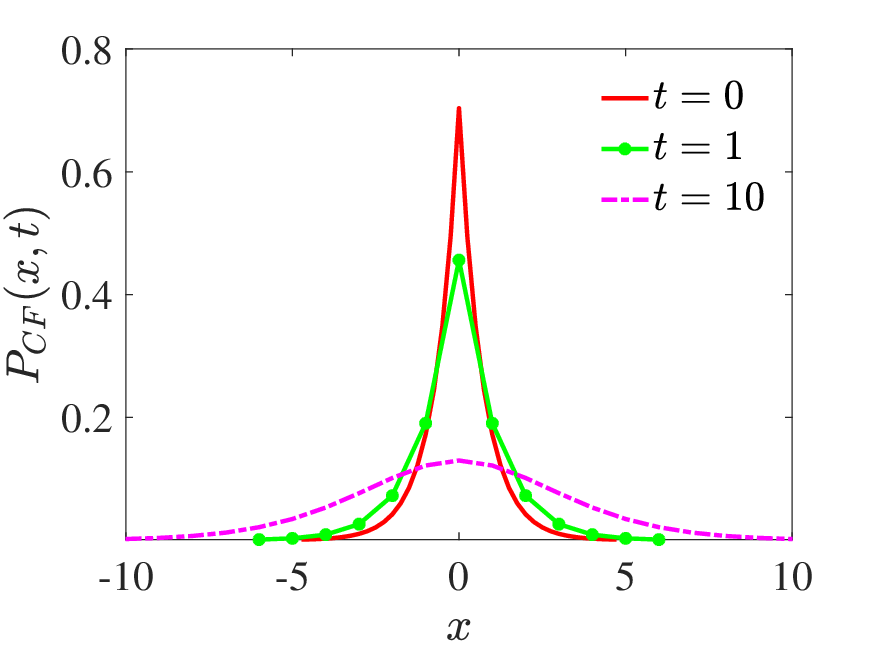}
\includegraphics[width=7.2cm]{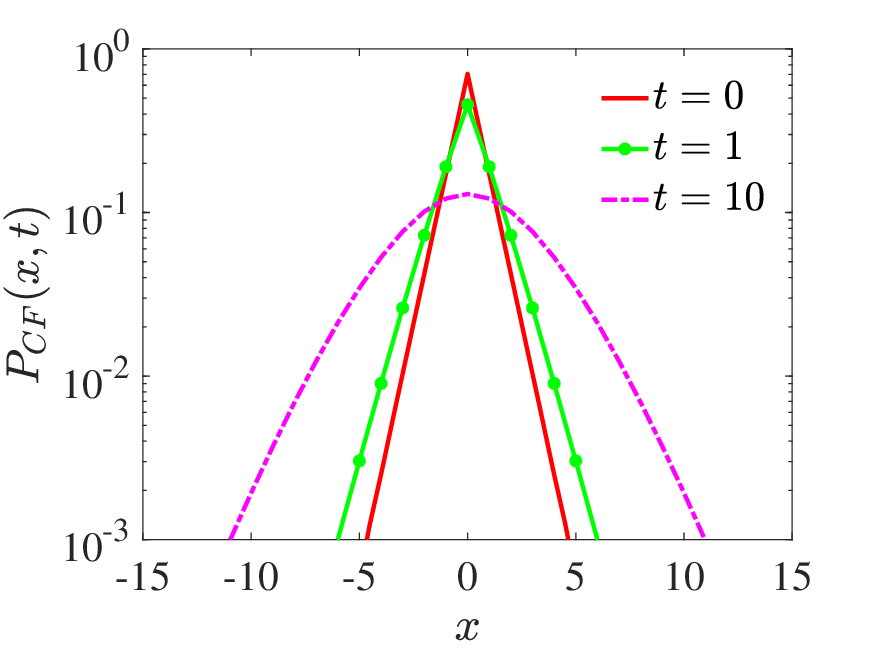}\\
(c)\includegraphics[width=7.2cm]{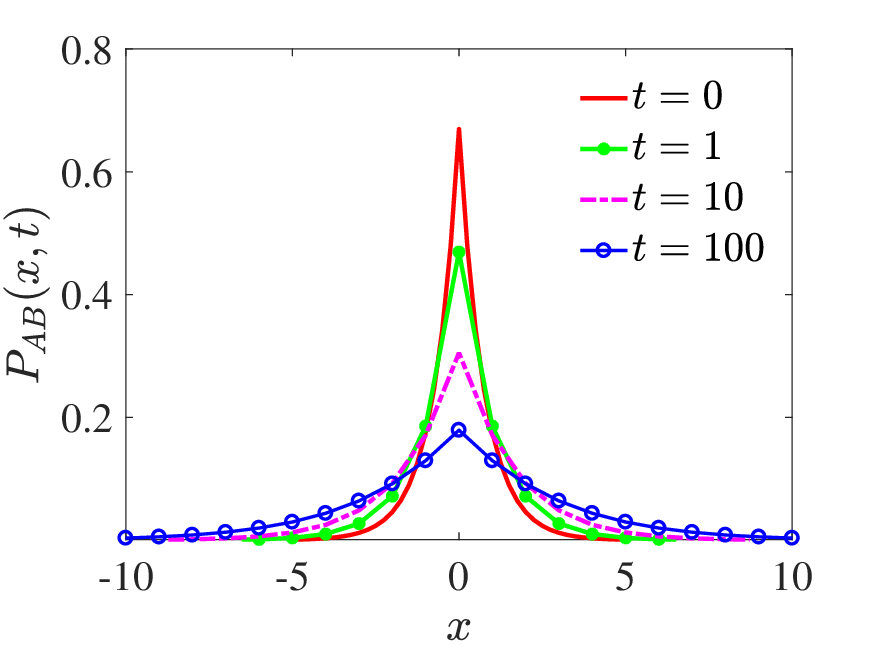}
\includegraphics[width=7.2cm]{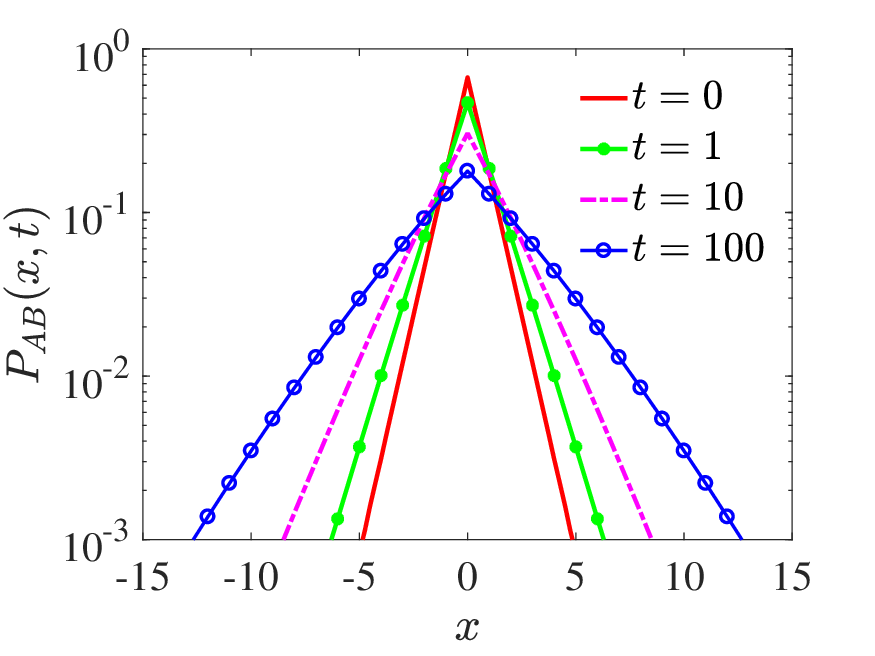}
(d)\includegraphics[width=7.2cm]{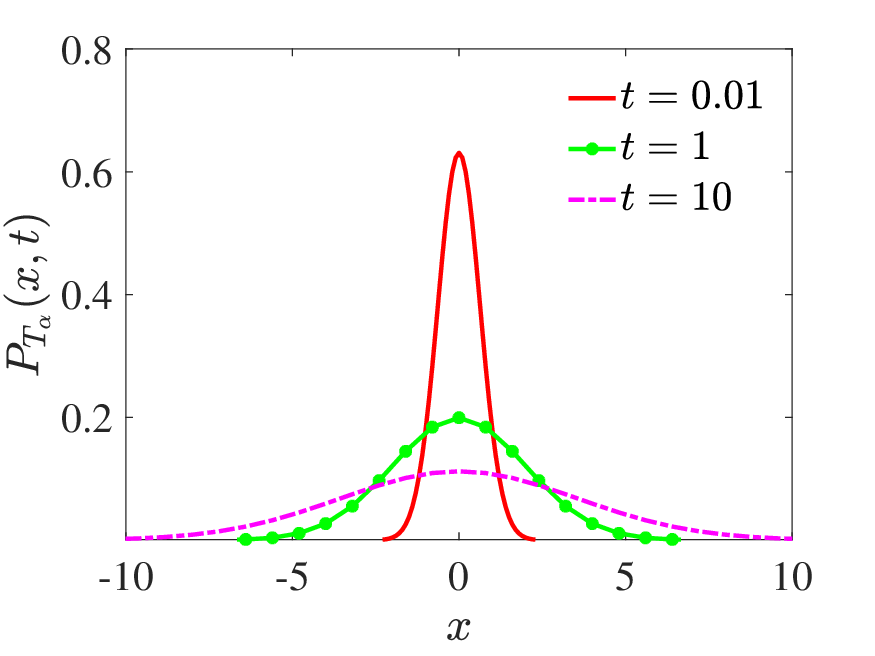}
\includegraphics[width=7.2cm]{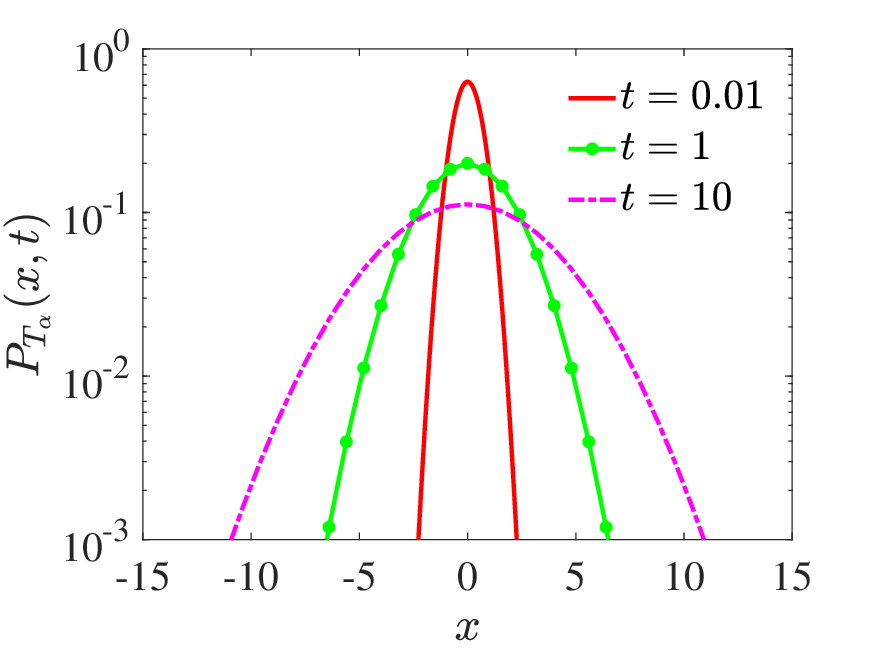}
\caption{PDF of diffusion equations with initial condition $P(x,0)=\delta(x)$
with different integro-differential
operators for $K_{\alpha}=1$, $\tau=1$, and $\alpha=0.5$: (a) Caputo-fractional,
(b) Caputo-Fabrizio, (c) Atangana-Baleanu, and (d) Conformable. The solutions
for the Caputo-FDE, CF-GDE and AB-GDE are obtained by applying an inverse
Laplace transform to Eq.~(\ref{lapsol}). Note that both PDFs for the CF- and
AB-GDE have a Laplace shape with finite width at $t=0$. The solution
(\ref{pdf-diff-conf-eq1}) of the conformable diffusion equation is Gaussian at
all times. The shapes are produced using Matlab code.}
\label{Fig1}
\end{figure*}

We digress briefly to shed some light on the delicate issue of placing
specific forms of exponential tempering or other non-singular kernels in
the integro-differential operators used above.
We consider the case of the CF-operator. Let us use the Schneider-Wyss
idea and start with the integral form (\ref{intde}) of the diffusion equation.
We naively replace the integral on the right hand side with the operator
(note that an analogous choice was made in Ref.~\cite{lenzi})
\begin{equation}
_0^{CF}D_t^{-\alpha}=\frac{\tau^{\alpha-1}}{1-\alpha}\int_0^tf(t')\exp\left(
-\frac{\alpha(t-t')}{(1-\alpha)\tau}\right)dt',
\end{equation}
which is an integral operator with exponential tempering similar to the
formulation of the CF-operator (\ref{CFdef}). Instead of the solution
(\ref{eq3}), for the initial condition $P_0(x)=\delta(x)$ we then obtain
\begin{equation}
P(k,s)=\frac{[\alpha/[(1-\alpha)\tau]+s]/s}{\alpha/[(1-\alpha)\tau]+s+K_
{\alpha}\tau^{\alpha-1}k^2}.
\end{equation}
The normalization is fulfilled, as for $k=0$ the Laplace transform reads
$1/s$. The initial condition can be found from this equation by setting
$s\to\infty$, which is $1/s$ to leading order. Inverse Fourier transform
indeed reproduces the initial condition $P_0(x)=\delta(x)$.

The second moment is obtained by differentiation twice with respect to $k$
and setting $k=0$,
\begin{equation}
\langle x^2(t)\rangle=\frac{2K_{\alpha}\tau^{\alpha}(1-\alpha)}{\alpha}\left(
1-\exp\left(-\frac{\alpha t}{(1-\alpha)\tau}\right)\right).
\end{equation}
At short times we find normal diffusion,
\begin{equation}
\langle x^2(t)\rangle\sim2K_{\alpha}\tau^{\alpha-1}t
\end{equation}
with the effective diffusion coefficient $K_{\alpha}\tau^{\alpha-1}$, and at
long times the saturation value
\begin{equation}
\langle x^2(t)\rangle\sim\frac{2(1-\alpha)K_{\alpha}\tau^{\alpha}}{\alpha}
\end{equation}
is reached. This convergence to a stationary plateau is similar to what was
observed previously for the tempering of FBM, that effected a behavior
consistent with confinement \cite{daniel}.

\begin{figure}
\includegraphics[width=7.2cm]{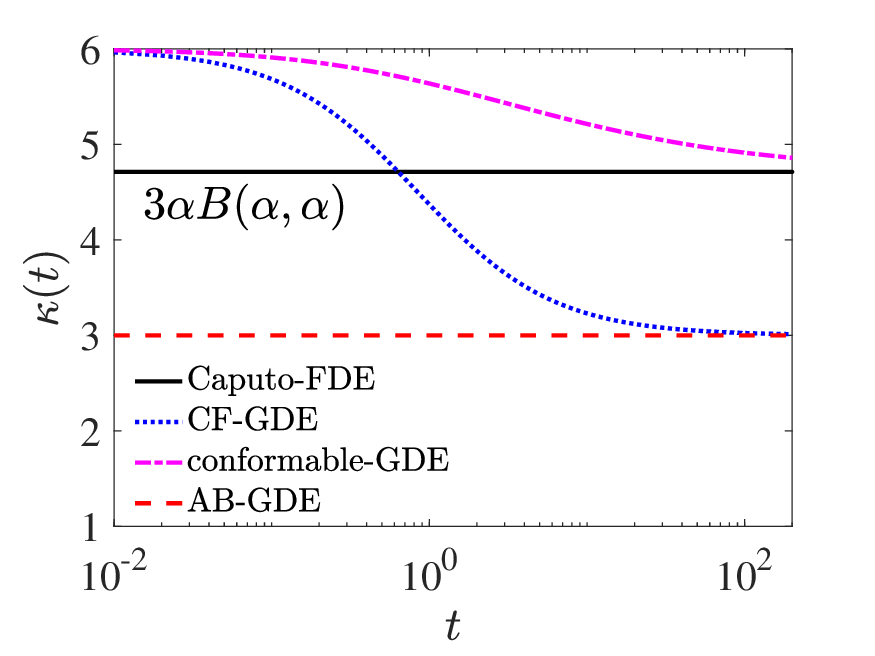}
\caption{Kurtosis $\kappa$ for the GDE with different integro-differential
operators for $K_{\alpha}=1$, $\tau=1$, and $\alpha=0.5$.}
\label{Fig2}
\end{figure}

\begin{figure}
\includegraphics[width=7.2cm]{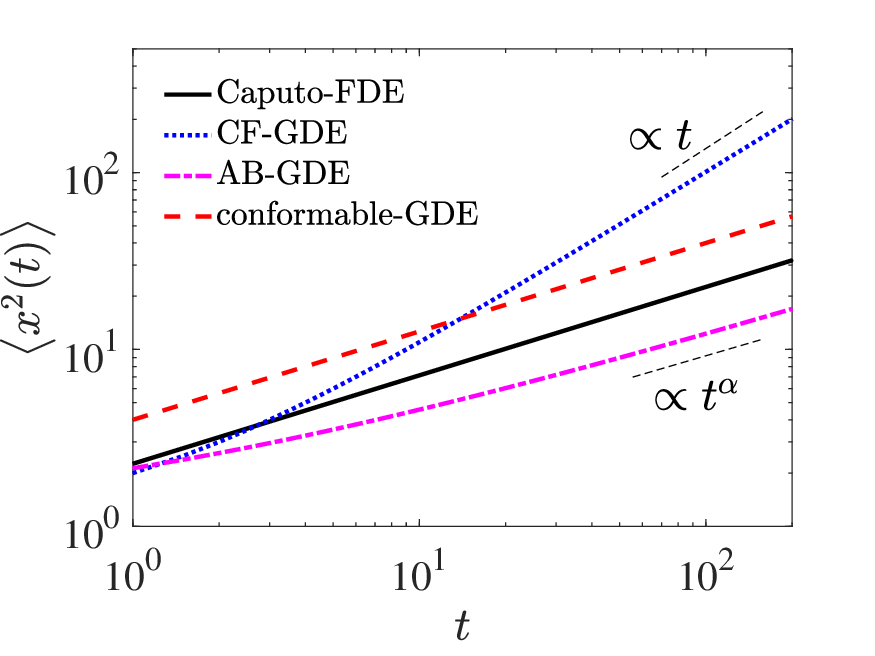}
\caption{MSD for the GDEs with different integro-differential operators
for $K_{\alpha}=1$, $\tau=1$, and $\alpha=0.5$.}
\label{Fig20}
\end{figure}
The inverse Fourier transform of $P(k,s)$ becomes
\begin{eqnarray}
\nonumber
P(x,s)&=&\frac{\sqrt{\pi\{\alpha/[(1-\alpha)\tau]+s\}/2}}{s\sqrt{K_{\alpha}
\tau^{\alpha-1}}}\\
&&\times\exp\left(-\frac{|x|\sqrt{\alpha/[(1-\alpha)\tau]+s}}{\sqrt{K_{\alpha}
\tau^{\alpha-1}}}\right).
\end{eqnarray}
At $s\to\infty$,
\begin{eqnarray}
\nonumber
P(x,s)&\sim&\frac{1}{s}\lim_{s\to\infty}\sqrt{\frac{\pi/2}{K_{\alpha}\tau^{\alpha
-1}/s}}\exp\left(-\frac{|x|\sqrt{s}}{\sqrt{K_{\alpha}\tau^{\alpha-1}}}\right)\\
&\sim&\frac{1}{s}\delta(x),
\end{eqnarray}
where we used the limiting form of the $\delta$-function. Thus, indeed, this
form leads back to the consistent initial value. In the opposite long-time
limit corresponding to $s\to0$, we have
\begin{equation}
P(x,s)\sim\frac{1}{s}\sqrt{\frac{\pi/2}{K_{\alpha}\tau^{\alpha}(1-\alpha)/
\alpha}}\exp\left(-\frac{|x|\sqrt{\alpha/(1-\alpha)}}{\sqrt{K_{\alpha}\tau
^{\alpha}}}\right).
\end{equation}
We thus have the stationary limit
\begin{equation}
P(x)\sim\sqrt{\frac{\pi/2}{K_{\alpha}\tau^{\alpha}(1-\alpha)/\alpha}}\exp\left(
-\frac{|x|\sqrt{\alpha/(1-\alpha)}}{\sqrt{K_{\alpha}\tau^{\alpha}}}\right).
\end{equation}
This is a Laplace distribution. We showed here for a simple modification of
the integral form of the diffusion equation using a CF-type exponential
tempering in the integral, how the time evolution of the PDF $P(x,t)$ will
reach a non-trivial stationary value. Physically, this can be viewed as a
consequence of introducing a finite time scale by the exponential factor in
the integral. Beyond this time the contributions of the integral to the
dynamics of $P(x,t)$ become exponentially small.

\subsubsection{SBM and conformable diffusion equation}

SBM in the absence of an external force, $F(x)=0$ in Eq.~(\ref{FPESBM}), and
for the standard initial condition $P(x,0)=\delta(x)$ has the Gaussian shape
\cite{lim2002self,jeon2014scaled},
\begin{eqnarray}
P(x,t)=\frac{1}{\sqrt{4\pi K_{\alpha}t^{\alpha}}}\exp\left(-\frac{x^2}{4K_
{\alpha}t^{\alpha}}\right),
\end{eqnarray}
with the associated MSD
\begin{eqnarray}
\label{SBMMSD}
\langle x^2(t)\rangle=2K_{\alpha}t^{\alpha}.
\end{eqnarray}

Concurrently, the GDE based on the conformable derivative is
\begin{eqnarray}
\label{diff-conf-eq1}
T_{\alpha}P(x,t)=K_{\alpha}\frac{\partial^2}{\partial x^2}P(x,t).
\end{eqnarray}
For $P(x,0)=\delta(x)$ and after applying the conformable Laplace transform
together with a Fourier transform
\begin{equation}
\label{fourier}
\mathscr{F}\{g(x)\}\equiv\int_{-\infty}^{\infty}g(x)\exp(ikx)dx,
\end{equation}
we find
\begin{equation}
\hat{\tilde{P}}_{T_\alpha}(k,s)=\frac{1}{s+K_{\alpha}k^2}.
\end{equation}
This directly leads to the PDF in $(x,t)$-space,
\begin{equation}
\label{pdf-diff-conf-eq1}
P_{T_\alpha}(x,t)=\sqrt{\frac{\alpha}{4\pi K_{\alpha}t^{\alpha}}}\exp\left(
-\frac{\alpha x^2}{4K_{\alpha}t^{\alpha}}\right).
\end{equation}
The Gaussianity of this PDF implies that the kurtosis is $\kappa_{T_{\alpha}}
(t)=3$. The MSD encoded by the PDF (\ref{diff-conf-eq1}) reads
\begin{eqnarray}
\langle x^2(t)\rangle_{T_\alpha}=\frac{2K_\alpha}{\alpha}t^\alpha.
\end{eqnarray}
Thus, the conformable-GDE has the same PDF, MSD, and kurtosis as subdiffusive
SBM. The PDF (\ref{pdf-diff-conf-eq1}) of the conformable-GDE is displayed in
Fig.~\ref{Fig1} (d). The associated MSD and kurtosis are shown in
Figs.~\ref{Fig2} and \ref{Fig20}. We will now show that the addition of a
constant force (constant drift) allows one to distinguish between these two
models.

\subsection{Generalized Fokker-Planck equation with drift}
\label{gfpe}

From Sec.~\ref{subsection3B} we conclude that the CF- and AB-operators in the
formulation of the GDE (\ref{ffpe}) (for $F(x)=0$) do not provide a consistent
formulation. We therefore do not consider them further here. We therefore limit
our discussion of anomalous diffusion equation with drift to the Caputo, SBM, and conformable formulations.

\subsubsection{Caputo diffusion equation with drift}

\begin{figure*}
(a)\includegraphics[width=0.43\textwidth]{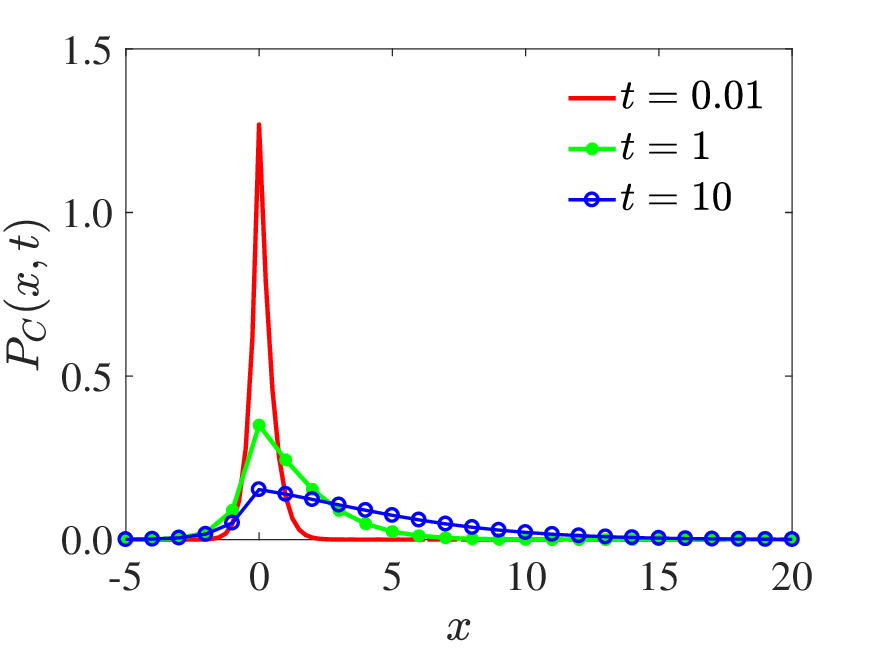}
\includegraphics[width=0.43\textwidth]{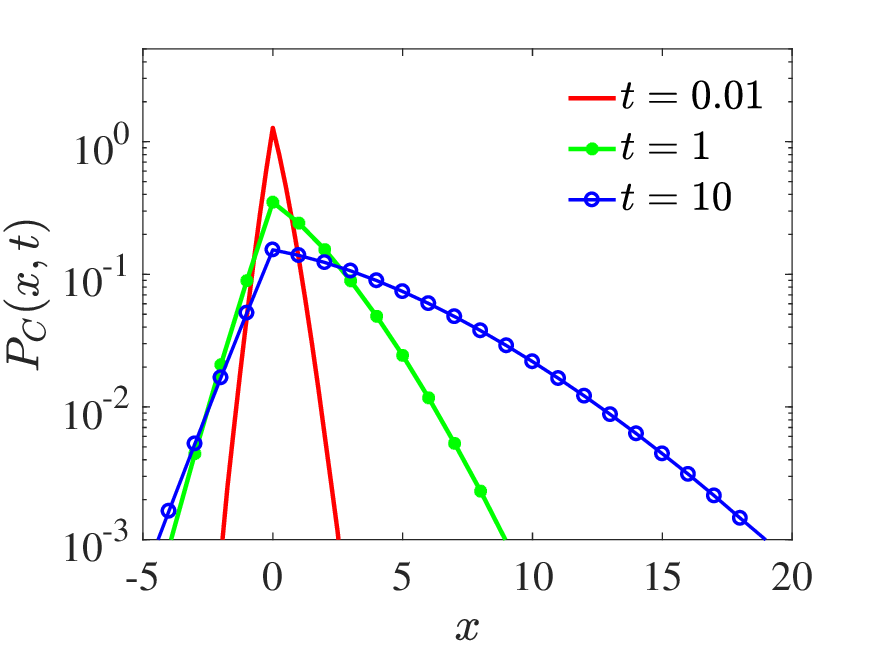}
(b)\includegraphics[width=0.43\textwidth]{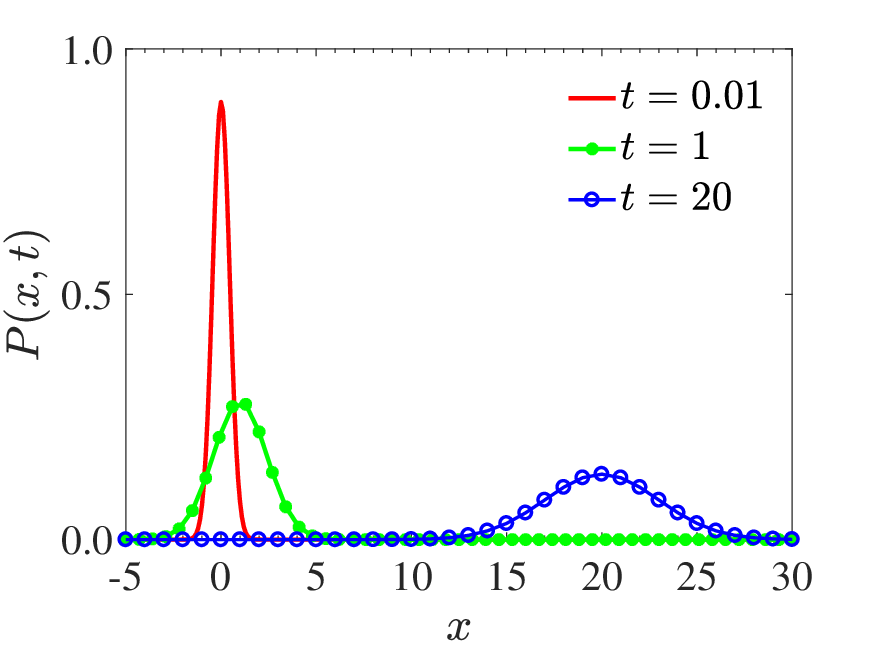}
\includegraphics[width=0.43\textwidth]{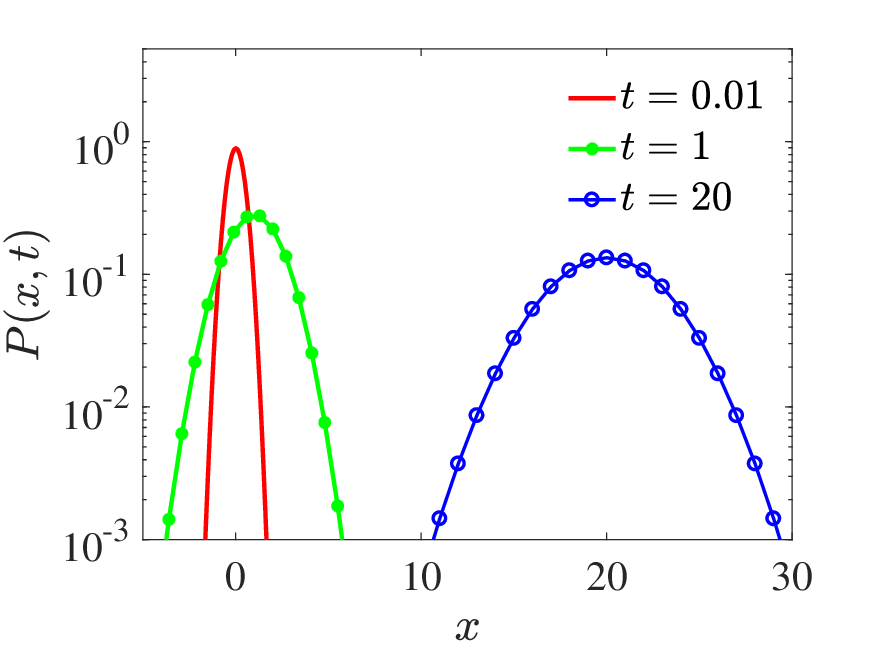}
(c)\includegraphics[width=0.43\textwidth]{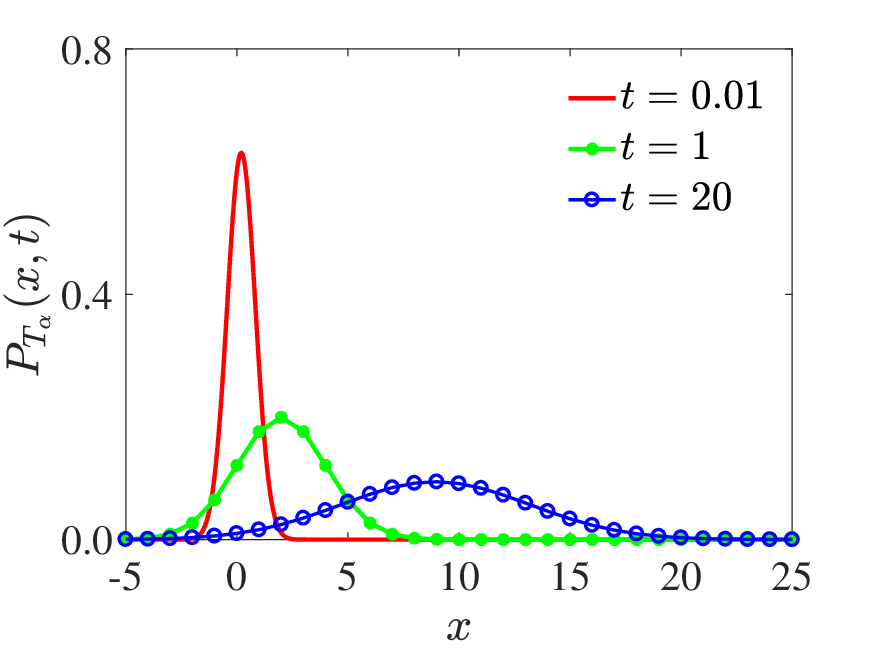}
\includegraphics[width=0.43\textwidth]{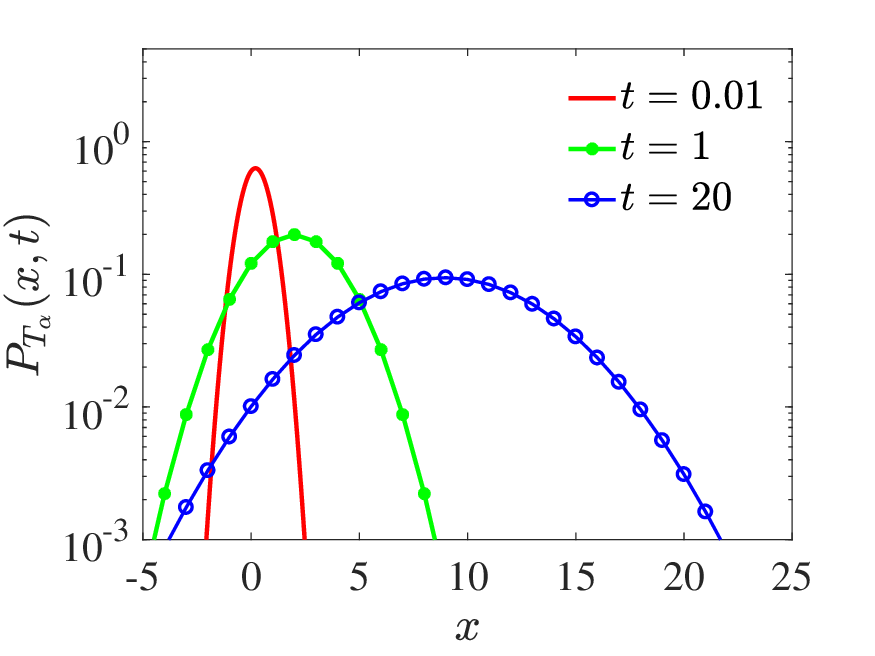}
\caption{PDFs for drift-GDEs with initial condition $P(x,0)=\delta(x)$ of (a)
Caputo, (b) SBM, (c) and conformable forms.
We use the parameters $\alpha=0.5$, $v=v_{\alpha}=1$, and $K_{\alpha}=1$. Note
that the position of the maximum of the PDF for the Caputo-FDE is stationary in
space, corresponding to particles that have not moved up to time $t$. The maximum
of the PDF for SBM with drift moves faster than that for the conformable-GDE.}
\label{Fig4}
\end{figure*}

The Caputo Fokker-Planck equation with drift reads (see also Ref.~\cite{dieterich2015fluctuation})
\begin{equation}
\label{CADE}
^C_0D^{\alpha}_t[P(x,t)]=K_{\alpha}\frac{\partial^2}{\partial x^2}P(x,t)
-v_{\alpha}\frac{\partial}{\partial x}P(x,t).
\end{equation}
From the CTRW perspective, Eq.~(\ref{CADE}) appears as the diffusion (long
time and space) limit of a walk with unequal probabilities to jump to the
right and to the left \cite{mebakla1,igor_pre,teza}. In Laplace
space the solution of Eq.~(\ref{CADE}) for the initial condition $P_0(x)=
\delta(x)$ becomes
\begin{eqnarray}
\nonumber
\tilde{P}_C(x,s)&=&\frac{s^{\alpha-1}}{\sqrt{v_{\alpha}^2+4K_\alpha s^\alpha}}\\
&\times&\exp\left(\frac{v_{\alpha}x}{2K_\alpha}-|x|\frac{\sqrt{v_{\alpha}^2
+4K_\alpha s^\alpha}}{2K_\alpha}\right).
\label{CADEL}
\end{eqnarray}
Applying a numerical inverse Laplace transformation to Eq.~(\ref{CADEL}), we
show this PDF in Fig.~\ref{Fig4}a.

The first moment and second moment encoded by the FDE (\ref{CADE}) are
\begin{equation}
\label{cfde1}
\langle x(t)\rangle_C=\frac{v_{\alpha}}{\Gamma(\alpha+1)}t^{\alpha}
\end{equation}
and
\begin{equation}
\langle x^2(t)\rangle_C=\frac{2K_{\alpha}}{\Gamma(\alpha+1)}t^{\alpha}
+\frac{2v_{\alpha}^2}{\Gamma(2\alpha+1)}t^{2\alpha}.
\label{cfde2}
\end{equation}
The MSD can then be calculated as
\begin{eqnarray}
\nonumber
\langle(\Delta x)^2\rangle_C&=&\frac{2K_{\alpha}}{\Gamma(\alpha+1)}t^{\alpha}\\
&+&\left[\frac{2}{\Gamma(2\alpha+1)}-\frac{1}{\Gamma(\alpha+1)^{2}}\right]
v^2_{\alpha}t^{2\alpha}.
\label{deltaDAC}
\end{eqnarray}
As well known from CTRW with a drift \cite{igor_pre,scher,report} the MSD
contains a term proportional to $v^2_{\alpha}t^{2\alpha}$, i.e., for $1/2<
\alpha<1$ an effective superdiffusion occurs. This is due to the strong
separation of particles stuck at the origin from mobile, advected particles.

\subsubsection{SBM with drift and conformable diffusion equation with drift}

The Fokker-Planck equation for SBM with drift is
\begin{equation}
\frac{\partial}{\partial t}P(x,t)=\mathscr{K}_{\alpha}(t)\frac{\partial^2}{
\partial x^2}P(x,t)-v\frac{\partial}{\partial x}P(x,t),
\end{equation}
from which we obtain the PDF
\begin{equation}
\label{eq30}
P(x,t)=\frac{1}{\sqrt{4\pi K_{\alpha}t^{\alpha}}}\exp\left(-\frac{(x-vt)^2}
{4K_{\alpha}t^{\alpha}}\right),
\end{equation}
which is shown in Fig.~\ref{Fig4}b. The associated moments are
\begin{equation}
\langle x(t)\rangle=vt,
\end{equation}
\begin{equation}
\langle x^2(t)\rangle=2K_{\alpha}t^{\alpha}+v^2t^2,
\end{equation}
such that the MSD is
\begin{equation}
\langle(\Delta x)^2\rangle=2K_{\alpha}t^{\alpha}.
\end{equation}
In contrast to the Caputo-FDE case, for SBM the Galilean invariance is preserved,
giving rise to the similarity variable $x-vt$ in the PDF (\ref{eq30}).

The GDE with drift based on the conformable derivative has the form
\begin{equation}
\label{eq23}
T_{\alpha}P(x,t)=K_{\alpha}\frac{\partial^2}{\partial x^2}P(x,t)-v_{\alpha}
\frac{\partial}{\partial x}P(x,t).
\end{equation}
Application of the conformable Laplace transform and Fourier transform yields
\begin{equation}
\label{eq25}
\hat{\tilde{P}}_{T_{\alpha}}(k,s)=\frac{1}{s+(K_{\alpha}k^2+iv_{\alpha}k)}.
\end{equation}
From this form we obtain the PDF in $(x,t)$-space,
\begin{equation}\label{eq27}
P_{T_{\alpha}}(x,t)=\sqrt{\frac{\alpha}{4\pi K_{\alpha}t^{\alpha}}}\exp
\left(-\frac{\alpha(x-v_{\alpha}t^{\alpha}/\alpha)^2}{4K_{\alpha}t^{\alpha}}
\right),
\end{equation}
which is shown in Fig.~\ref{Fig4}c.

\begin{figure}
(a)\includegraphics[width=7.2cm]{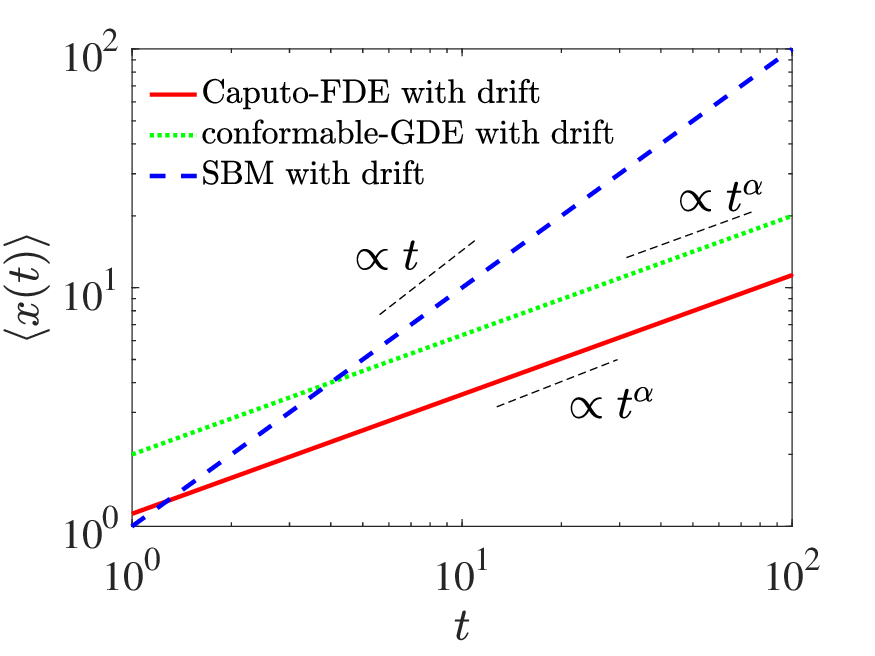}
(b)\includegraphics[width=7.2cm]{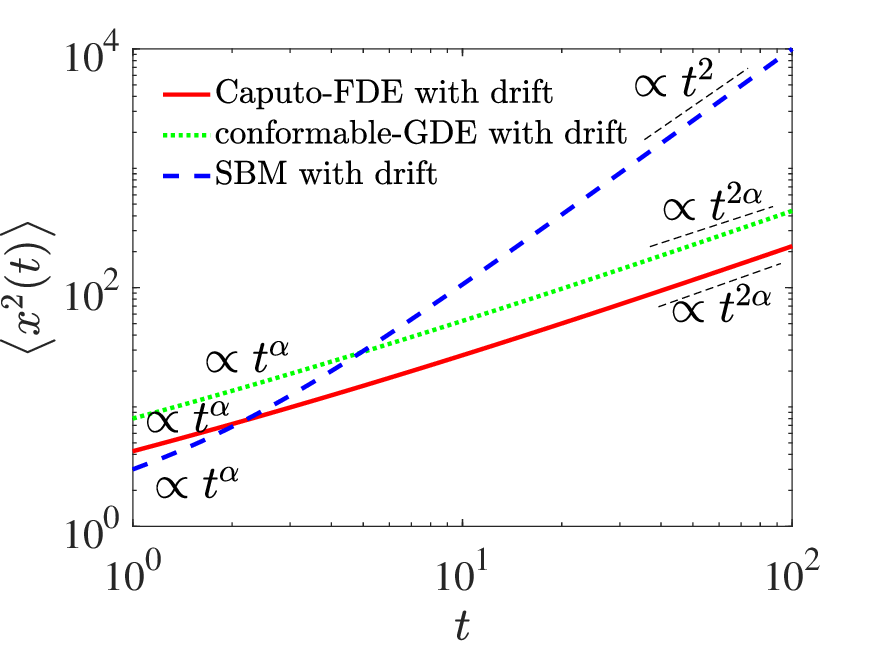}
(c)\includegraphics[width=7.2cm]{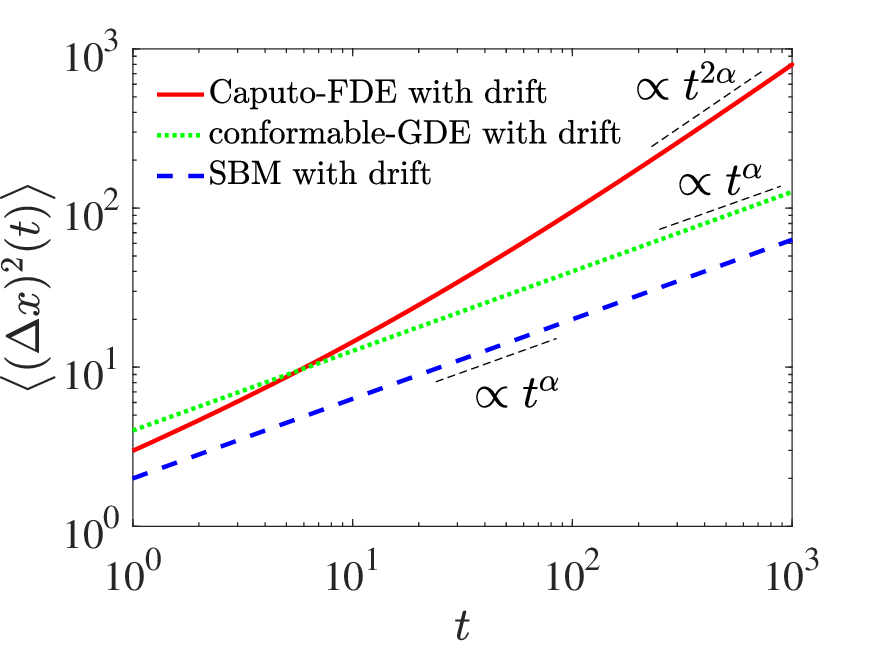}
\caption{First (a) and (b) second moments, and (c) MSD for Caputo, conformable,
and SBM forms. We chose $\alpha=0.5$, $v=v_{\alpha}=1$, and $K_{\alpha}=1$.}
\label{Fig7}
\end{figure}

The associated first moment is
\begin{equation}
\label{eq28}
\langle x(t)\rangle=\frac{v_{\alpha}t^{\alpha}}{\alpha},
\end{equation}
and the second moment reads
\begin{equation}
\label{eq29}
\langle x^2(t)\rangle=\frac{2K_{\alpha}}{\alpha}t^{\alpha}+\frac{v_{\alpha}^2}
{\alpha^2}t^{2\alpha}.
\end{equation}
The MSD is then
\begin{equation}
\label{VCDED}
\langle(\Delta x)^2\rangle=\frac{2K_{\alpha}}{\alpha}t^{\alpha}.
\end{equation}

The behaviors of the moments for the Caputo, SBM, and conformable forms of the
generalized motion are shown in Fig.~\ref{Fig7}, from which one can distinguish
SBM with drift from the Caputo and conformable drift-GDEs via the first moment:
for SBM the first moment grows linearly in $t$, while for the other two cases
a scaling with $t^{\alpha}$ occurs. From the MSD one can then distinguish the
Caputo and conformable forms, respectively scaling as $t^{2\alpha}$ and $t^{
\alpha}$ in the long time limit.

\section{Generalized Langevin equations}
\label{section4}

We now turn to generalizations of the stochastic formulation of diffusive
processes based on the Langevin equation, see \cite{daniel,
liemert2017generalized,deng2009ergodic,kou2004generalized,porra1996generalized}
for details.

\subsection{Generalized Langevin equations in the
force-free case}

In the absence of an external force, $F(x)=0$, we consider formulations with
different Caputo-, CF-, and AB-operators, SBM \cite{jeon2014scaled}, and FBM
\cite{jeon2010fractional,mandelbrot1968fractional}.

\subsubsection{Caputo and non-singular integro-differential operators}
\label{section4A1}

Applying the Caputo, CF- and AB-operators, we have the generalized Langevin
equation,
\begin{equation}
\label{caputoLE}
\frac{d^{\alpha}}{dt^{\alpha}}x(t)=\sqrt{2K_{\alpha}}\xi(t),
\end{equation}
for which we impose the initial condition $x(0)=0$.
Here $d^{\alpha}/dt^{\alpha}$ represents the Caputo, CF- and AB-operators.
Formal integration produces
\begin{equation}
x(t)=\sqrt{2K_{\alpha}}\int_0^tH(t-t')\xi(t')dt',
\end{equation}
where the integral kernel $H$ stands for
\begin{eqnarray}
\nonumber
H_C(t)&=&\frac{t^{\alpha-1}}{\Gamma(\alpha)},\\
\nonumber
H_{CF}(t)&=&(1-\alpha)\tau^{\alpha}\delta(t)+\alpha \tau^{\alpha-1},\\
H_{AB}(t)&=&(1-\alpha)\tau^{\alpha}\delta(t)+\alpha\frac{t^{\alpha-1}}{
\Gamma(\alpha)},
\label{hdef}
\end{eqnarray}
for the Caputo-, CF- and AB-operators, respectively.

\begin{figure*}
(a)\includegraphics[width=7.2cm]{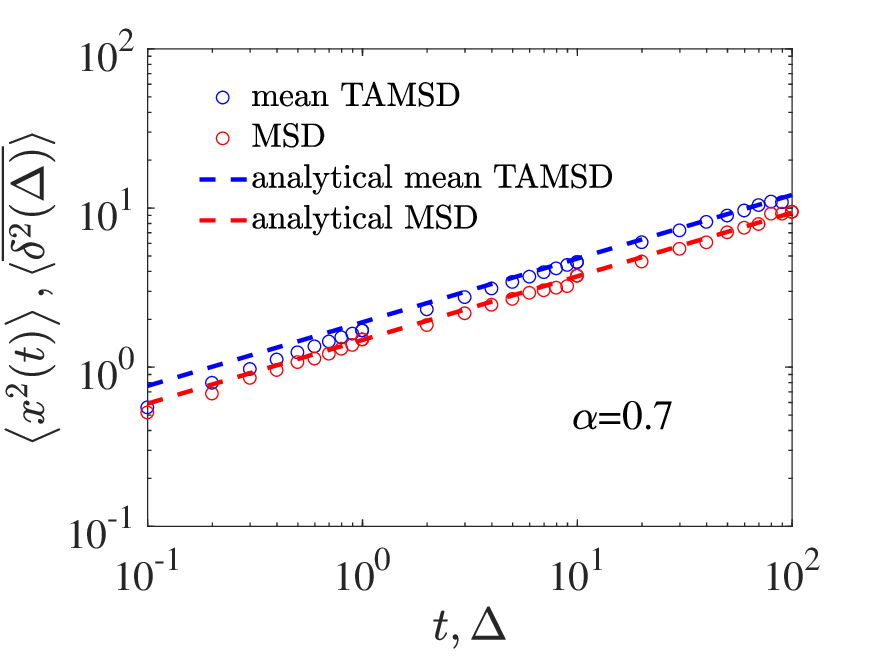}
\includegraphics[width=7.2cm]{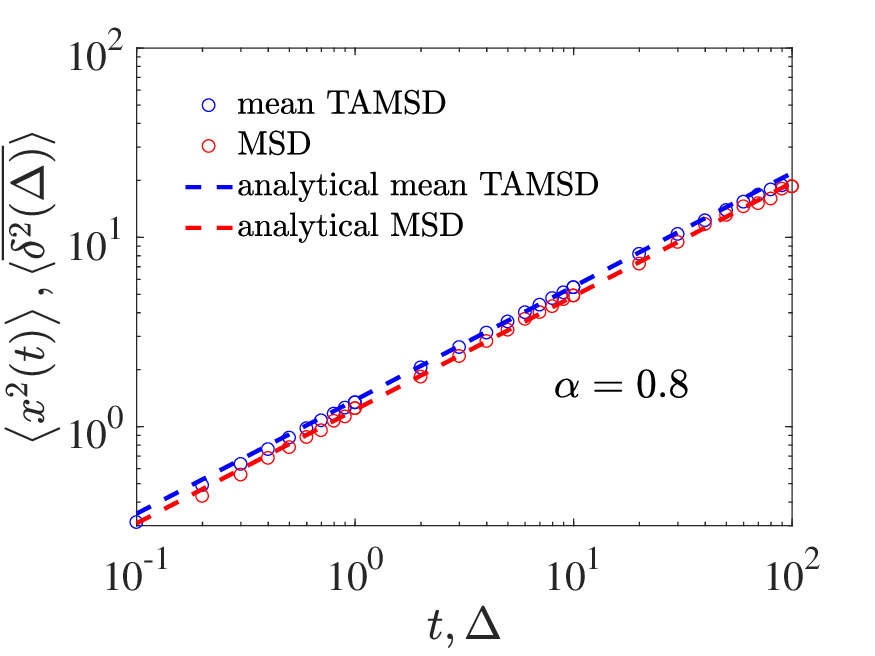}\\
(b)\includegraphics[width=7.2cm]{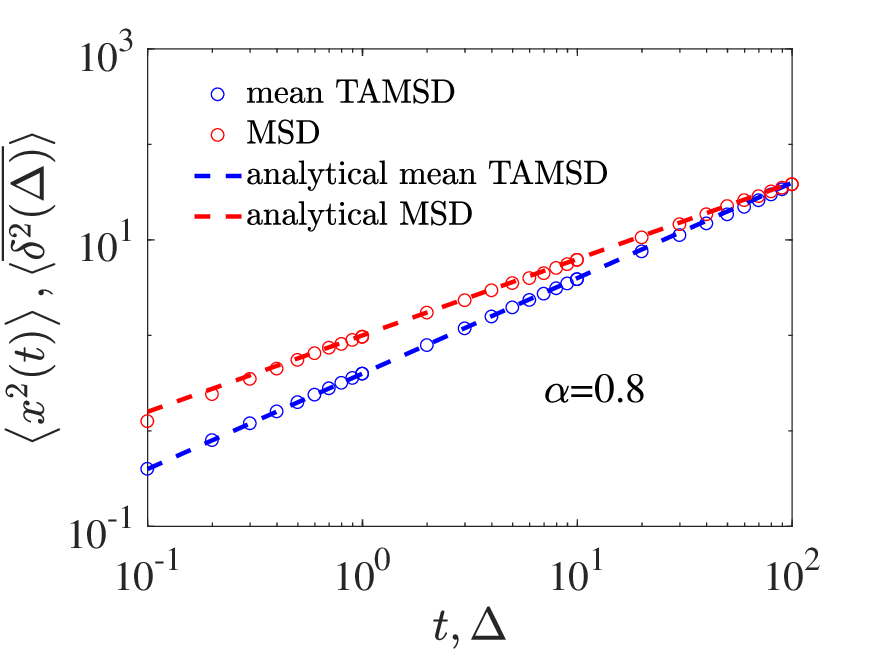}
\includegraphics[width=7.2cm]{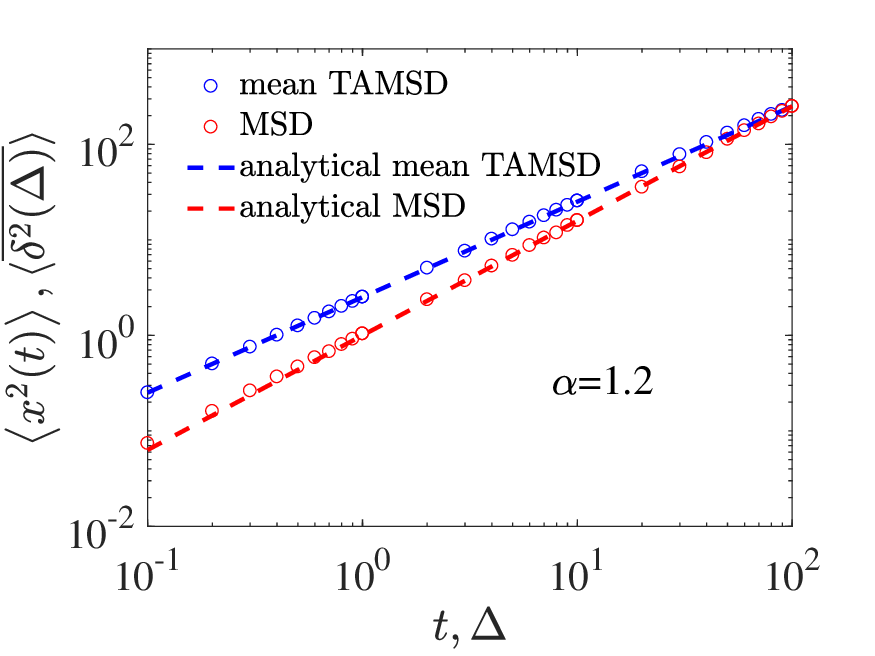}\\
(c)\includegraphics[width=7.2cm]{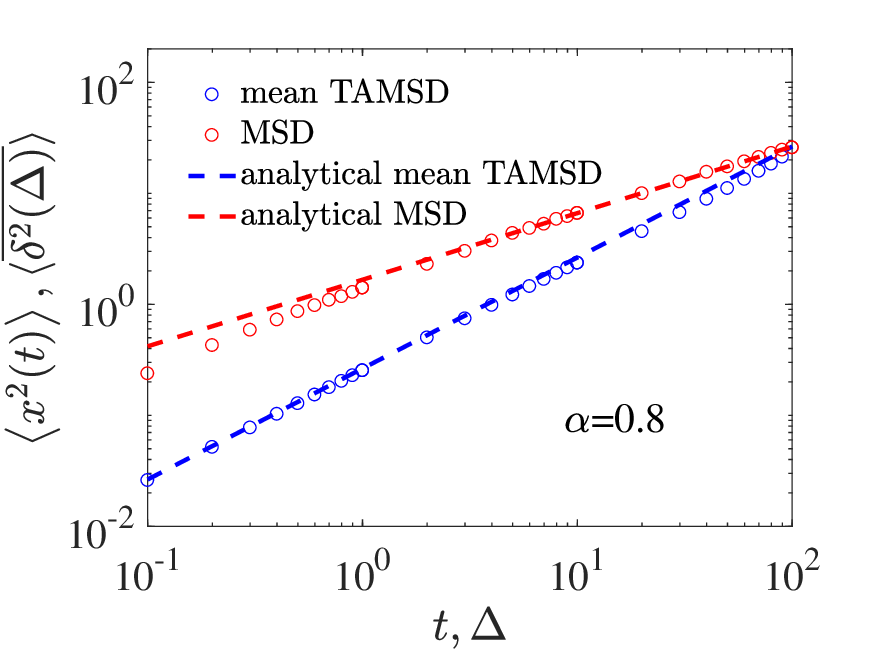}
\includegraphics[width=7.2cm]{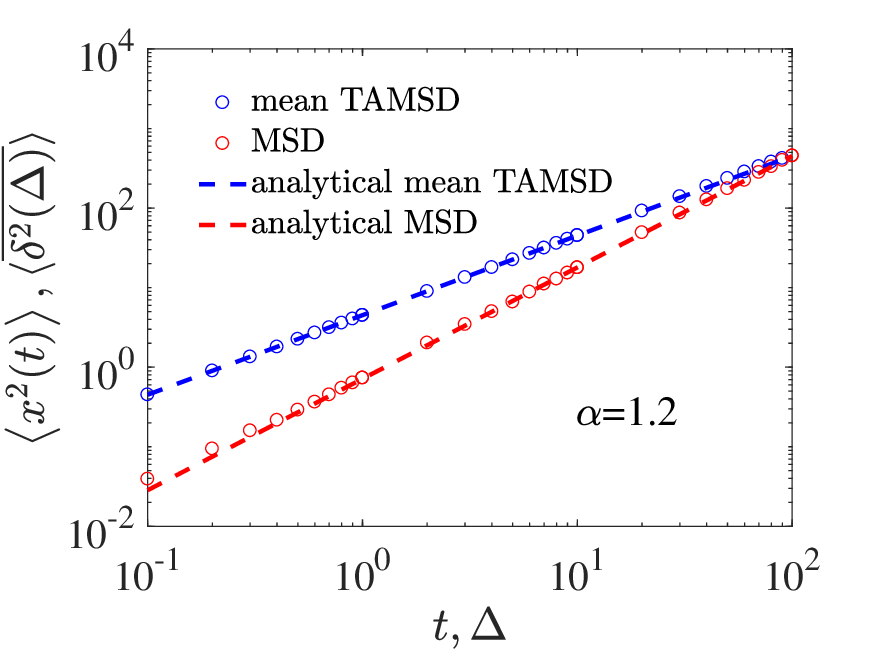}
\caption{Simulations and analytical solutions for MSD and mean TAMSD for the
(a) Caputo-, (b) SBM-, and (c) conformable-generalized Langevin equations,
for $K_{\alpha}=0.5$.}
\label{SLEMSD}
\end{figure*}

For the Caputo-fractional Langevin equation, the two-point correlation function
can be obtained in the form
\begin{equation}
\label{CaputoLEtwo}
\langle x(t_1)x(t_2)\rangle=\frac{2K_{\alpha}t_2^{\alpha}t_1^{\alpha-1}}{
\alpha\Gamma{(\alpha)}^2}\,_2F_1\left(1-\alpha,1;\alpha+1;\frac{t_2}{t_1}
\right),
\end{equation}
where we assume (without limitation of generality) that $t_1\ge t_2$, and where
\begin{equation}
_2F_1(a,b;c;z)=\frac{\Gamma(c)}{\Gamma(b)\Gamma(c-b)}\int_0^1\frac{t^{b-1}
(1-t)^{c-b-1}}{(1-tz)^a}dt
\end{equation}
is the hypergeometric function (see Eq.~(\ref{A14}) in App~\ref{AppendixB2}
for details of the derivation of Eq.~(\ref{CaputoLEtwo})). From
Eq.~(\ref{CaputoLEtwo}), the MSD follows for $t_1=t_2=t$, yielding
\begin{equation}
\label{CLE-MSD}
\langle x^2(t)\rangle=\frac{2K_{\alpha}}{(2\alpha-1)\Gamma{(\alpha)}^2}t^{
2\alpha-1},
\end{equation}
for $\alpha>1/2$ \cite{REM}. We also obtain the mean TAMSD (see
App~\ref{AppendixC1} for details) in the limit $\Delta/T\ll1$,
\begin{equation}
\label{CaputoTAMSD}
\left<\overline{\delta^2(\Delta)}\right>\sim\frac{2K_{\alpha}}{\Gamma(2\alpha)
\left|\cos(\pi\alpha)\right|}\Delta^{2\alpha-1}.
\end{equation}
While the MSD and the mean TAMSD for the Caputo-Langevin equation
thus have the same scaling exponent $2\alpha-1$, the two expressions have
different prefactors. That implies that the process encoded by the
Caputo-Langevin equation are non-ergodic in the Birkhoff-Boltzmann
sense. In the notation of \cite{aljaz}, we call such a case ultraweak
ergodicity breaking. Results of stochastic simulations for the MSD and the mean
TAMSD for the Caputo-Langevin equation are shown in Fig.~\ref{SLEMSD}a,
and we find good agreement with the theoretical results. Details on the
discrete simulations scheme for the different operators are provided in
App~\ref{AppendixA}. 

The fact that the Caputo-Langevin equation produces a non-stationary
dynamics can be anticipated from the autocovariance function (\ref{CaputoLEtwo}),
see \cite{stas} for further discussion. We finally report the displacement ACVF
of this process, which asymptotically reads $C_{\Delta}(t)\sim[2(\alpha-1)K_
{\alpha}/\{\alpha\Gamma^2(\alpha)\}]\Delta^{\alpha-1}t^{\alpha-2}$,
see App \ref{AppendixC2} for the derivation. We note that this behavior is
different from the Caputo-FDE based on CTRW processes, for which the ACVF
is zero beyond $t=\Delta$ \cite{stas}.

For the case of the CF- and AB-Langevin equations we focus on their
second moment,
\begin{equation}
\langle x^2(t)\rangle=2K_{\alpha}\int_0^tH^2(t')dt',
\end{equation}
where for the kernel $H$ the respective forms from Eq.~(\ref{hdef}) should be
substituted. Noticing that $\int_0^t\delta(t')^2dt'=\infty$, this means that
the second moment for both CF- and AB-formulations diverges. Details of the
derivations can be found in App~\ref{AppendixB2}. Similar to our observations
in the case of the FDE, the formulations in terms of the CF- and AB-operators
lead to inconsistent results, and we will not pursue these operators further.

\subsubsection{SBM- and conformable-generalized Langevin equation}
\label{subsection4B}

The formal solution of the SBM-Langevin equation (\ref{LESBM}) when $F(x)=0$
is
\begin{eqnarray}
x(t)=\sqrt{2\alpha K_\alpha}\int_0^t(t')^{\frac{\alpha-1}{2}}\xi(t')dt',
\end{eqnarray}
and the two-point correlation function reads
\begin{equation}
\label{tpc}
\langle x(t_1)x(t_2)\rangle=2K_{\alpha}[\min\{t_1,t_2\}]^{\alpha}.
\end{equation}
The MSD then has the power-law form \cite{jeon2014scaled}
\begin{eqnarray}
\label{SBMmsd}
\langle x^2(t)\rangle=2K_\alpha t^\alpha.
\end{eqnarray}
The mean TAMSD grows as \cite{jeon2014scaled}
\begin{equation}
\left<\overline{\delta^2(\Delta)}\right>=\frac{2K_\alpha\left[T^{\alpha+1}
-\Delta^{\alpha+1}-(T-\Delta)^{\alpha+1}\right]}{(\alpha+1)(T-\Delta)}.
\end{equation}
In the limit $\Delta/T\ll1$,
\begin{equation}
\label{sbmtamsd}
\left<\overline{\delta^2(\Delta)}\right>\sim2K_\alpha\Delta T^{\alpha-1},
\end{equation}
which is linear in the lag time $\Delta$. SBM is thus weakly non-ergodic in
the above Birkhoff-Boltzmann sense \cite{barkai2012single}. In contrast to
the ultraweak situation above, here the mean TAMSD explicitly depends on the
measurement time $T$. Simulations results for the MSD and the mean TAMSD for
SBM are shown in Fig.~\ref{SLEMSD}b.

The conformable Langevin equation
\begin{equation}
\label{eq20}
T_{\alpha}x(t)=\sqrt{2K_{\alpha}}\xi(t),
\end{equation}
can be rephrased by using the relation $T_{\alpha}[f(t)]=t^{1-\alpha}df(t)/dt$
between the conformable derivative and the first order derivative,
\begin{equation}
\frac{d}{dt}x(t)=\sqrt{2K_{\alpha}}t^{\alpha-1}\xi(t),
\end{equation}
and thus
\begin{equation}
\label{eq21}
x(t)=\sqrt{2K_{\alpha}}\int_0^t(t')^{\alpha-1}\xi(t')dt'.
\end{equation}
The two-point correlation function is
\begin{equation}
\label{CLECOR}
\langle x(t_1)x(t_2)\rangle=\frac{2K_{\alpha}}{2\alpha-1}[\min\{t_1,t_2\}]^{2
\alpha-1}.
\end{equation}
for $\alpha>1/2$. Finally, the MSD becomes
\begin{equation}
\label{CLEMSD}
\langle x^2(t)\rangle=\frac{2K_{\alpha}}{2\alpha-1}t^{2\alpha-1}.
\end{equation}
For the mean TAMSD we obtain
\begin{equation}
\left<\overline{\delta^2(\Delta)}\right>=\frac{C_1}{2\alpha}\frac{T^{2\alpha}}{
T-\Delta}\left[1-\left(\frac{\Delta}{T}\right)^{2\alpha}-\left(1-\frac{\Delta}{T}
\right)^{2\alpha}\right],
\end{equation}
where $C_1=2K_{\alpha}/[(2\alpha-1)\Gamma^2(\alpha)].$
In the limit $\Delta/T\ll1$,
\begin{equation}
\label{conftamsd}
\left<\overline{\delta^2(\Delta)}\right>\sim\frac{K_\alpha}{2\alpha-1}\Delta
T^{2\alpha-2}.
\end{equation}

The conformable-Langevin equation encodes a weakly non-ergodic and
non-stationary dynamic. Simulations results of the MSD and mean TAMSD for
the conformable-Langevin equation (\ref{eq20}) with different $\alpha$ are
shown in Fig.~\ref{SLEMSD}c. From Eqs.~(\ref{tpc}) and (\ref{CLECOR}) it
follows that both processes, SBM and conformable-Langevin equation motion have
independent increments, and thus the ACVFs defined in Eq.~(\ref{acvfdef})
vanishes for these processes, $C_{\Delta}(t)=0$. The analysis of the MSDs
(\ref{SBMmsd}) and (\ref{CLEMSD}) for the two processes shows that they have
the same time-scaling if we take the exponent $\alpha$ for SBM equal to the
exponent $2\alpha-1$ for the conformable-Langevin equation. A similar
conclusion follows from Eqs.~(\ref{sbmtamsd}) and (\ref{conftamsd}). Therefore,
the information encoded in the MSD and mean TAMSD is insufficient to distinguish
between these two processes. Analogously to the situation considered in Section
\ref{gfpe} for the SBM- and conformable-diffusion equations we will show that
adding a constant force (constant drift) allows us to distinguish between these
two Langevin models.

\subsubsection{FBM}
\label{subsection4C}

The formal solution of FBM for $F(x)=0$ is (\ref{LEFBM})
\begin{equation}
x(t)=\int_0^t\xi_{\alpha}(t')dt',
\end{equation}
with the two-point correlation
\begin{equation}
\langle x(t_1)x(t_2)\rangle=K_{\alpha}\left(t_1^{\alpha}+t_2^{\alpha}-\left|
t_1-t_2\right|^{\alpha}\right).
\end{equation}
Thus, the MSD has the power-law form
\begin{equation}
\label{FBMMSD}
\langle x^2(t)\rangle=2K_{\alpha}t^{\alpha}.
\end{equation}
The mean TAMSD of FBM is \cite{deng2009ergodic}
\begin{equation}
\left<\overline{\delta^2(\Delta)}\right>=2K_{\alpha}\Delta^{\alpha}.
\end{equation}
We conclude that this free FBM is ergodic and stationary. Finally, the
normalized autocovariance of FBM can be represented as $C_{\Delta}(t)/C_
{\Delta}(0)=[(t+\Delta)^{\alpha}-2t^{\alpha}+|t-\Delta|^{\alpha}/[2\Delta^
{\alpha}]$, for $\Delta\neq0$ \cite{jae}.

\subsection{Generalized Langevin equations with drift }

\subsubsection{Caputo-fractional Langevin equation with drift}

The Caputo-fractional Langevin equation with drift reads
\begin{equation}
^C_0D^{\alpha}_tx(t)=\sqrt{2K_{\alpha}}\xi(t)+v_{\alpha}.
\end{equation}
After a Laplace transformation,
\begin{equation}
\tilde{x}(s)=\frac{v_{\alpha}}{s^{1+\alpha}}+\sqrt{2K_{\alpha}}\frac{\xi(s)}{
s^{\alpha}}.
\end{equation}
Back-transforming to the time domain,
\begin{equation}
\label{CaputoLED}
x(t)=\frac{v_{\alpha}t^{\alpha}}{\Gamma(1+\alpha)}+\sqrt{2K_{\alpha}}
\int_0^t\frac{{(t-t')}^{\alpha-1}}{\Gamma(\alpha)}\xi(t')dt'.
\end{equation}
The first moment is then given by
\begin{equation}
\langle x(t)\rangle=\frac{v_{\alpha}t^{\alpha}}{\Gamma(1+\alpha)},
\end{equation}
which coincides with the result (\ref{cfde1}) of the Caputo-FDE. Similar to
the derivation of Eq.~(\ref{CaputoLEtwo}), the two-point correlation function
of the Caputo-fractional Langevin equation in the presence of a drift is
\begin{eqnarray}
\nonumber
\langle x(t_1)x(t_2)\rangle&=&\frac{v_{\alpha}^2}{\Gamma^2(1+\alpha)}(t_1t_2)^{
\alpha}+\frac{2K_{\alpha}t_2^{\alpha}t_1^{\alpha-1}}{\alpha\Gamma{(\alpha)}^2}\\
&&\times\,_2F_1\left(1-\alpha,1,\alpha+1,\frac{t_2}{t_1}\right),
\end{eqnarray}
where we assumed that $t_1<t_2$. The second moment is
\begin{equation}
\label{cflesec}
\langle x^2(t)\rangle=\frac{v_{\alpha}^2}{\Gamma^2(1+\alpha)}t^{2\alpha}+
\frac{2K_{\alpha}}{(2\alpha-1)\Gamma^2(\alpha)}t^{2\alpha-1}, \alpha>\frac{1}{2}.
\end{equation}
for $\alpha>1/2$. We finally obtain the MSD
\begin{equation}
\label{cflemsd}
\langle (\Delta x)^2(t)\rangle=\frac{2K_{\alpha}}{(2\alpha-1)\Gamma^2
(\alpha)}t^{2\alpha-1}.
\end{equation}
The forms of the second moment (\ref{cflesec}) and the MSD (\ref{cflemsd})
are different from their counterparts (\ref{cfde2}) and (\ref{deltaDAC})
for the Caputo-FDE. In the Caputo-fractional Langevin equation the drift
enters additively and is not affected by the memory in the fractional
operator. In contrast, for the Caputo-FDE the particles are immobilized
during the waiting times represented by the fractional operator.

\subsubsection{SBM- and conformable-Langevin equations with drift}

The SBM-Langevin equation with drift has the form
\begin{equation}
\frac{d}{dt}x(t)=\sqrt{2K_{\alpha}(t)}\xi(t)+v.
\end{equation}
The moments are readily calculated, yielding
\begin{equation}
\label{fm}
\langle x(t)\rangle=vt
\end{equation}
and
\begin{equation}
\langle x^2(t)\rangle=2K_{\alpha}t^{\alpha}+v^2t^2.
\end{equation}
Thus the MSD is
\begin{equation}
\langle(\Delta x)^2\rangle=2K_{\alpha}t^{\alpha}.
\end{equation}

The conformable-Langevin equation with drift is
\begin{equation}
\label{eq31}
T_{\alpha}x(t)=\sqrt{2K_{\alpha}}\xi(t)+v_{\alpha}.
\end{equation}
With the relation $T_{\alpha}f(t)=t^{1-\alpha}df(t)/dt$, we obtain
\begin{equation}
\frac{d}{dt}x(t)=t^{\alpha-1}\left(\sqrt{2K_{\alpha}}\xi(t)+v_{\alpha}\right),
\end{equation}
and thus
\begin{eqnarray}
\nonumber
x(t)&=&\int_0^t(t')^{\alpha-1}\left(\sqrt{2K_{\alpha}}\xi(t')+v_{\alpha}\right)
dt'\\
&=&\frac{v_{\alpha}}{\alpha}t^{\alpha}+\sqrt{2K_{\alpha}}\int_0^t(t')^{\alpha
-1}\xi(t')dt',
\label{eq32}
\end{eqnarray}
and thus the first moment is equivalent to the deterministic form
\begin{equation}
\label{eq33}
\langle x(t)\rangle=\frac{v_{\alpha}}{\alpha}t^{\alpha}.
\end{equation}
The two-point correlation function becomes
\begin{equation}
\langle x(t_1)x(t_2)\rangle=\frac{v_{\alpha}^2}{\alpha^2}(t_1t_2)^{\alpha}
+\frac{2K_{\alpha}}{2\alpha-1}t_2^{2\alpha-1}.
\end{equation}
valid for $\alpha>1/2$. The second moment follows as
\begin{equation}
\langle x^2(t)\rangle=\frac{v_{\alpha}^2}{\alpha^2}t^{2\alpha}
+\frac{2K_{\alpha}}{2\alpha-1}t^{2\alpha-1}.
\end{equation}
Finally, the MSD has the $v_{\alpha}$-independent form
\begin{equation}
\label{VCLED}
\langle(\Delta x)^2(t)\rangle=\frac{2K_{\alpha}}{2\alpha-1}t^{2\alpha-1}
\end{equation}
for $\alpha>1/2$.
One can see from Eqs.~(\ref{fm}) and (\ref{eq33}) that the response to a constant
force is different for SBM-Langevin and conformable-Langevin equations, thus
allowing us to distinguish between these two anomalous diffusion models.

\subsubsection{FBM with drift}

We finally consider the FBM Langevin equation with drift,
\begin{equation}
x(t)=\int_0^t dt'\xi_{\alpha}(t')+vt,
\end{equation}
such that
\begin{equation}
\langle x(t)\rangle=vt.
\end{equation}
The two-point correlation behaves as
\begin{equation}
\langle x(t_1)x(t_2)\rangle=K_{\alpha}\left(t_1^{\alpha}+t_2^{\alpha}-|t_1
-t_2|^{\alpha}\right)+v^2t_1t_2.
\end{equation}
The second moment encoded in this form is
\begin{equation}
\langle x^2(t)\rangle=2K_{\alpha}t^{\alpha}+v^2t^2.
\end{equation}
Finally, the MSD reads
\begin{equation}
\langle(\Delta x)^2(t)\rangle=2K_{\alpha}t^{\alpha}.
\end{equation}

\begin{figure}
(a)\includegraphics[width=7.2cm]{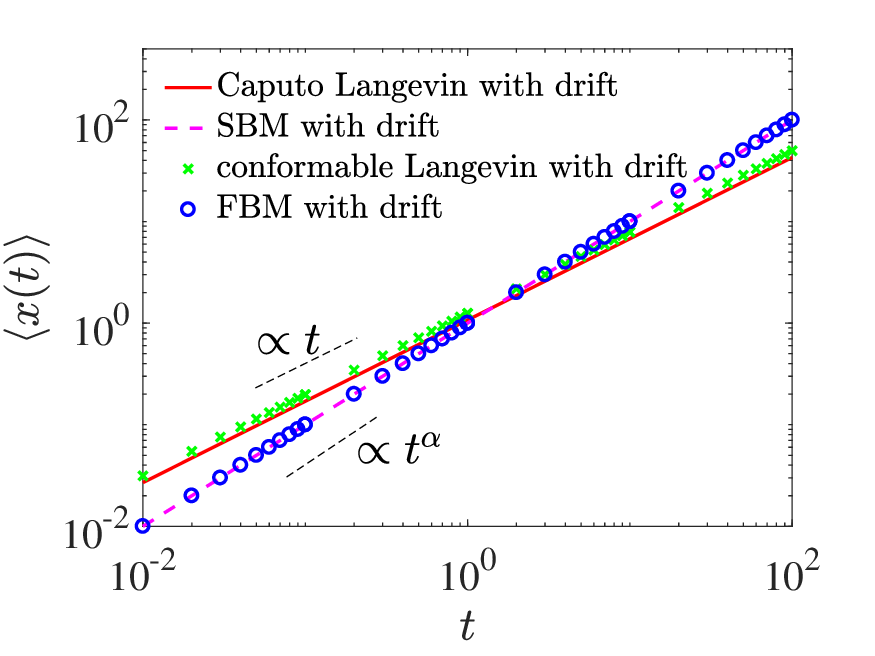}
(b)\includegraphics[width=7.2cm]{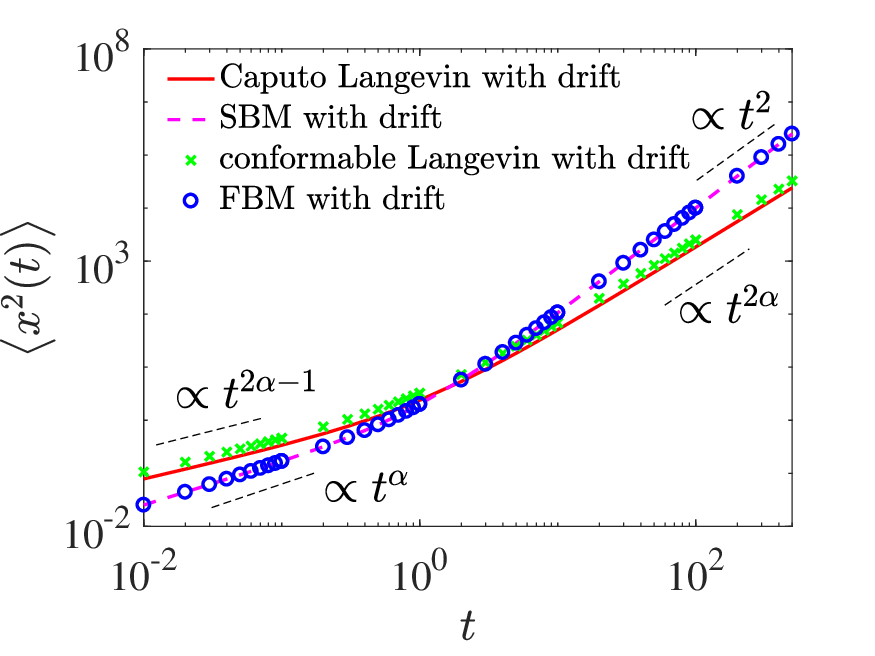}
(c)\includegraphics[width=7.2cm]{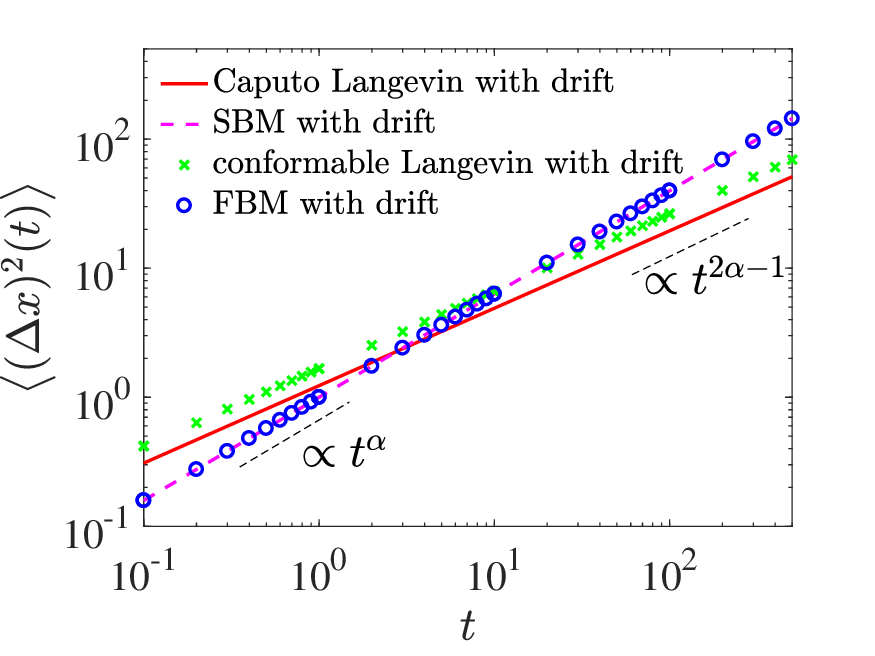}
\caption{First (a) and second (b) moments and (c) MSD for Caputo-fractional,
conformable-, SBM-, and FBM-Langevin equations with drift for $v=1$, $\alpha
=0.8$, and $K_{\alpha}=0.5$. Note that the results of SBM and FBM fully coincide in all three panels.}
\label{LEDmoment}
\end{figure}

The moments for the different processes are shown in Fig.~\ref{LEDmoment}. From
the above discussion we see that the first moments of the Caputo-fractional and
and conformable-Langevin equations with drift have a power-law form, which is
distinct from the linear time dependence in the SBM- and FBM-Langevin equations.

\begin{table*}
\begin{ruledtabular}
\begin{tabular}{ccccc}
Fokker-Planck Eq. && Caputo $(0<\alpha\leq 1)$ & conformable $(0<\alpha\leq 1)$ &
SBM $(\alpha>0)$ \\ \hline
{$F(x)=0$ } & {$P(x,t)$} & {$\frac{1}{2\sqrt{K_{\alpha}t^{\alpha}}} M_{\frac
{\alpha}{2}}\left(\frac{|x|}{\sqrt{K_{\alpha} t^{\alpha}}} \right)$}
& {$\sqrt{\frac{\alpha}{4\pi K_{\alpha}t^{\alpha}}}
\exp\left(-\frac{\alpha x^{2}}{4K_{\alpha}t^{\alpha}}\right)$} &
{$\frac{1}{\sqrt{4\pi K_{\alpha}t^{\alpha}}}\exp\left(-\frac{x^{2}}{
4K_{\alpha}t^{\alpha}}\right)$} \\
& $\langle x^2(t)\rangle$ & $\frac{2K_{\alpha}}{\Gamma(\alpha+1)}t^{\alpha}$
& $\frac{2K_{\alpha}}{\alpha}t^{\alpha}$ & $2K_{\alpha}t^{\alpha}$ \\\hline
$F(x)=v$ & $P(x,t)$ & $\mathscr{L}^{-1}[\tilde{P}_C(x,s)]$\footnote{$\tilde{P}
_C(x,s)=\frac{s^{\alpha-1}}{\sqrt{v_{\alpha}^2+4K_\alpha s^\alpha}}\exp\left(
\frac{v_{\alpha}x}{2K_\alpha}-|x|\frac{\sqrt{v_{\alpha}^2+4K_\alpha s^\alpha}}{
2K_\alpha}\right).$} &
{$\sqrt{\frac{\alpha}{4\pi K_{\alpha}t^{\alpha}}}\exp\left(-\frac{\alpha (x-
\frac{v_{\alpha}t^{\alpha}}{\alpha})^2}{4K_{\alpha}t^{\alpha}}\right)$} &
$\frac{1}{\sqrt{4\pi K_{\alpha}t^{\alpha}}}\exp\left(-\frac{(x-vt)^{2}}{
4K_{\alpha}t^{\alpha}}\right)$ \\
& $\langle x(t)\rangle$ & $\frac{v_{\alpha}}{\Gamma(\alpha+1)}t^{\alpha}$ &
$v_{\alpha}\frac{t^{\alpha}}{\alpha}$ & $vt$ \\
& $\langle x^2(t)\rangle$ & $\frac{2K_{\alpha}}{\Gamma(\alpha+1)}t^{\alpha}
+\frac{2v_{\alpha}^2}{\Gamma(2\alpha+1)}t^{2\alpha}$ & $\frac{2K_{\alpha}}{\alpha}
t^{\alpha}+\frac{v_{\alpha}^2}{\alpha^2}t^{2\alpha}$ & $2K_{\alpha}t^{\alpha}
+v^2t^2$ \\
& $\langle(\Delta x)^2(t)\rangle$ & $\frac{2K_{\alpha}}{\Gamma(\alpha+1)}
t^{\alpha}+at^{2\alpha}$\footnote{$a=\left(\frac{2}{\Gamma(2\alpha+1)}-\frac{1}{
\Gamma(\alpha+1)^2}\right)v_{\alpha}^2$.} & $\frac{2K_{\alpha}}{\Gamma(
\alpha+1)}t^{\alpha}$ & $2K_{\alpha}t^{\alpha}$\\
\end{tabular}
\end{ruledtabular}
\caption{\label{tab:table1}
Central results for the displacement PDFs and moments of the generalized
Fokker-Planck equations with Caputo-fractional and conformable derivatives
and SBM for the cases without and with drift.}
\end{table*}

\begin{table*}
\begin{ruledtabular}
\begin{tabular}{cccccc}
Langevin Eq. & & Caputo $(\frac{1}{2}<\alpha\leq 1)$ & conformable $(\frac{1}{2}<\alpha\leq 1)$ & SBM $(\alpha>0)$ & FBM $(0<\alpha<2)$\\ \hline
$F(x)=0$ &$\langle x^2(t)\rangle$ & $\frac{2K_\alpha}{(2\alpha-1)\Gamma(\alpha)
^2}t^{2\alpha-1}$ & $\frac{2K_\alpha}{2\alpha-1}t^{2\alpha-1}$ & $2K_{\alpha}
t^{\alpha}$ & $2K_{\alpha}t^{\alpha}$\\
& $\left<\overline{\delta^2(\Delta)}\right>$ & $\sim\frac{2K_{\alpha}}{\Gamma(
2\alpha)\left|\cos(\pi\alpha)\right|}\Delta^{2\alpha-1}$ & $\sim\frac{K_\alpha}{
2\alpha-1}\Delta T^{2\alpha-2}$ & $\sim2K_\alpha\Delta T^{\alpha-1}$ & $2K_{
\alpha}\Delta^{\alpha}$\\
& $C_{\Delta}(t),\,t\gg\Delta$ & $\simeq(\alpha-1)K_{\alpha}\Delta^{\alpha-1}t^{
\alpha-2}$ & $0$ & $0$ & $\simeq(\alpha-1)\Delta^{\alpha-2}$\\\hline
$F(x)=v,v_{\alpha}$ & $\langle x(t)\rangle $ & $v_{\alpha}\frac{t^\alpha}{
\Gamma(1+\alpha)}$ & $v_{\alpha}\frac{t^\alpha}{\alpha}$ & $vt$ & $vt$\\
& $\langle x^2(t)\rangle$ & $\frac{v_{\alpha}^2}{\Gamma(1+\alpha)^2}t^{2\alpha}
+\frac{2K_\alpha}{(2\alpha-1)\Gamma(\alpha)^2}t^{2 \alpha-1}$ & $\frac{v_{\alpha}
^2}{\alpha^2}t^{2\alpha}+\frac{2K_\alpha}{2\alpha-1}t^{2\alpha-1}$ &
$2K_{\alpha}t^{\alpha}+v^2t^2$ & $2K_{\alpha}t^{\alpha}+v^2t^2$ \\
& $\langle(\Delta x)^2(t)\rangle$ & $\frac{2K_\alpha}{(2\alpha-1)\Gamma(\alpha)
^2}t^{2\alpha-1}$ & $\frac{2K_\alpha}{2\alpha-1}t^{2\alpha-1}$ & $2K_{\alpha}
t^{\alpha}$ &$2K_{\alpha}t^{\alpha}$\\
\end{tabular}
\end{ruledtabular}
\caption{\label{tab:table2}
Central results for the moments, mean TAMSD, and autocovariance function of the
generalized Langevin equations with Caputo-fractional and conformable derivatives,
SBM and FBM, without and with drift.}
\end{table*}

\section{Conclusions}
\label{section6}

Fractional dynamic equations of the relaxation and diffusion types have been
used in science and engineering for considerable time. Traditionally, these
generalized equations were used in the Riemann-Liouville and Caputo types.
These are particularly suitable for the formulation of initial value problems
posed at $t=0$. Other formulations such as the Weyl-Riesz forms have been used,
as well, e.g., in the context of generalized rheological models for harmonic
driving \cite{helmut,helmut1}. More recently, additional definitions of
fractional and conformable operators have been proposed and discussed in
literature. We here studied generalized diffusion and Langevin equations,
comparing the classical Caputo-fractional forms with the CF-, AB-, and
conformable-generalized differential operators. We also compare these results to
two other anomalous diffusion processes, SBM and FBM.

In our analysis we find that the formulations in terms of the CF- and AB-operators
lead to inconsistent results for the PDFs and MSDs in both the GDE and generalized
Langevin equation cases. While this point requires further analysis from a more
mathematical point of view, we here did not pursue the formulations in terms of
these operators further. A possible solution for the incorrect incorporation of
the initial values for these two operators in the traditional formulation (note
that the integral formulation as outlined in the Introduction for the CF- and
AB-cases produces the same results) may be that instead of an initial condition
at $t=0$, the initial condition has to be formulated on an interval. We discussed
an alternative formulation similar to the CF-GDE in which the initial condition
is consistently incorporated. The latter formulation will deserve further
analysis in the future.

Results for the moments, the MSD, and the PDF of the different formulations
using Caputo-fractional, conformable-, SBM-, and FBM-dynamic equations are
summarized in Tabs.~\ref{tab:table1} and \ref{tab:table2}. Generally we see
that the first moments in the presence of drift are identical for both GDE and
Langevin formulations for each of the Caputo-, conformable-, and SBM-models,
while their higher order moments are different between GDE and Langevin
descriptions for the Caputo and conformable cases. The Caputo-, conformable-,
and SBM-cases exhibit non-stationarity and non-ergodic behavior. The PDFs are
Gaussian in all cases apart from the Caputo-FDE. We also see that by combining
moments in the presence and absence of a constant drift velocity the three
models can be distinguished.

Of particular interest here is the formulation in terms of the conformable
derivative. The resulting PDF turns out to be the same as the PDF for SBM
in the force-free case, after a renormalization of the generalized diffusion
coefficient. In the presence of a drift, both processes differ in the scaling
of the first moment and in the form the drift enters the PDF. Despite its
"local" definition, the conformable-GDE is weakly non-ergodic and shows
aging properties. These effects are visible in the comparison of the MSD with
the mean TAMSD as well as in the two-point correlation function in the
conformable-Langevin equation case.

We also note that we chose to present our analysis in dimensional units. This
requires the use of generalized diffusion coefficients and drift velocities.
Including dimensionality allows for an explicit extraction of the parameters
from measurements. It also demonstrates the different ways (local or with a
generalized exponent) the drift enters the different model dynamics.

Our study should help to assess and compare different formulations of
GDEs and generalized Langevin equations. A similar analysis should be performed
for the case of a harmonic confinement of the test particle. Moreover,
different forms of crossovers to normal dynamics should be studied.

\begin{acknowledgments}

We acknowledge support from the German Science Foundation (DFG grant ME
1535/12-1) and NSF-BMBF CRCNS (grant 2112862/STAXS). AVC acknowledges support
by the Polish National Agency for Academic Exchange (NAWA). We also acknowledge
the National Natural Science Foundation of China (52204110, 51827901, 52121003,
52142302, 51904309), the 111 Project (B14006) and the Yueqi Outstanding Scholar
Program of CUMTB (2017A03). This paper is also funded by the China Scholarship
council.

\end{acknowledgments}

\appendix

\section{Integral versions of CF- and AB-operators}\label{AppendixD}

To find the inverse operator of Eq.~(\ref{CFdef}), the CF-integral, we take
$0<\alpha\leq1$ and consider the equation
\begin{equation}
^{CF}_0D^{\alpha}_{t}f(t)=u(t).
\end{equation}
After a Laplace transform we obtain
\begin{eqnarray}
\nonumber
\mathscr{L}\{f(t)\}(s)&=&\frac{1}{s}f(0)+\frac{\alpha\tau^{\alpha-1}}{sM(
\alpha)}\mathscr{L}\{u(t)\}(s)\\
&&+\frac{(1-\alpha)\tau^{\alpha}}{M(\alpha)}\mathscr{L}\{u(t)\}(s).
\end{eqnarray}
Rearranging and after inverse Laplace transformation we deduce that
\begin{equation}
f(t)=\frac{(1-\alpha)\tau^{\alpha}}{M(\alpha)}u(t)+\frac{\alpha\tau^{\alpha
-1}}{M(\alpha)}\int_0^tu(s)ds+f(0).
\end{equation}
Thus, the CF-integral is defined as
\begin{equation}
_0^{CF}I_t^\alpha u(t)=\frac{(1-\alpha)\tau^{\alpha}}{M(\alpha)}u(t)+\frac{
\alpha\tau^{\alpha-1}}{M(\alpha)}\int_0^tu(s)ds.
\end{equation}

Similarly, the modified AB-integral corresponding to the operator (\ref{ABdef})
is
\begin{eqnarray}
\nonumber
_0^{AB}\mathrm{I}_t^\alpha f(t)&=&\frac{(1-\alpha)}{B(\alpha)}\tau^{\alpha}f(t)\\
&+&\frac{\alpha}{B(\alpha)\Gamma(\alpha)}\int_0^tf(t')(t-t')^{\alpha-1}dt',
\end{eqnarray}
for $0<\alpha\leq1$.

\section{Integro-differential operators in diffusion and Langevin equations}
\label{AppendixB}

\subsection{Integro-differential operators in diffusion equations}
\label{AppendixB1}

For the Caputo-FDE the PDF (\ref{C2}) can also be represented in terms of
the Fox $H$-function \cite{mathai}
\begin{equation}
\label{eq6}
P_C(x,t)=\frac{1}{\sqrt{4K_\alpha t^\alpha}}H_{1,1}^{1,0}\left[\frac{|x|}{
\sqrt{K_\alpha t^\alpha}}\left|\begin{array}{l}(1-\alpha/2,\alpha/2)\\
(0,1)\end{array}\right.\right].
\end{equation}

For the PDFs of the CF- and AB-GDEs, we check whether their PDFs in Laplace
space are completely monotonic \cite{Rschilling2010}. To this end we first
check the complete monotonicity of expressions (\ref{lapsol}), (\ref{eqs}),
and (\ref{eqs1}).
First, we introduce the completely monotone functions (CMF) and Bernstein
functions (BF), as well as some useful properties of these two types of
functions. CMFs can be represented as Laplace transforms of a non-negative
function $p(t)$, i.e., $m(x)=\int_0^{\infty}p(t)\exp{(-xt)}dt$. They are
defined on the non-negative half-axis and have the property that $(-1)^n
m^{(n)}(x)\geq0$ for all $n\in\mathbb{N}_0$ and $x\geq0$. The following
property holds true for CMFs \cite{Rschilling2010}:

(i) The product $m(x)=m_1(x)m_2(x)$ of two CMFs $m_1(x)$ and $m_2(x)$ is
again a CMF.

The Bernstein functions \cite{Rschilling2010} are non-negative functions, whose derivative is completely monotone. They have the property that $(-1)^{(n-1)}b^{
(n)}(x)\geq0$, for all $n=\mathbb{N}$. The Bernstein functions have the following
two properties:

(ii) A composition $b_1\big(b_2(s)\big)$ of Bernstein functions is again a
Bernstein function.

(iii) A composition $m\big(b(x)\big)$ of a CMF $m(x)$ and a Bernstein function
$b(x)$ is a CMF.

From these properties it follows that the function $\exp(-ub(x))$ is completely
monotone for $u>0$ if $b(x)$ is a Bernstein function.

Now let $f_1(s)=1/\sqrt{s(c_1s+c_2)}=1/[\sqrt{s}\sqrt{c_1s+c_2}]$. As
$1/\sqrt{s}$ is a CMF, $1/\sqrt{c_1s+c_2}$ is a CMF, so from property
(i) $f_1(s)$ is also a CMF. Let $f_2(s)=\exp(-\sqrt{s/(c_1s+c_2)}|x|)$.
As $\sqrt{s}$ is a BF and $s/(c_1s+c_2)$ is a BF, then from property (ii),
$\sqrt{s/(c_1s+c_2)}$ is a BF. As $\exp(-s)$ is a CMF, then from property
(iii), $f_2(s)$ is a CMF. Consequently
$\tilde{P}_{CF}(x,s)=f_1(s)f_2(s)$ is a CMF. Similarly, $\tilde{P}_{AB}(x,s)$
is a CMF. From the above we can ensure that the PDFs of the CF- and AB-GDEs
obtained from Eqs.~(\ref{lapsol}), (\ref{eqs}) and (\ref{eqs1}) represent
proper PDFs. 

Now we focus on the calculation of the second moment from the GDE (\ref{eqd}),
\begin{equation}
\langle x^2(t)\rangle=\int_{-\infty}^{\infty}x^2P(x,t)dx,
\end{equation}
with initial value $\langle x^2(0)\rangle=\int_{-\infty}^{\infty}x^2P(x,0)dx=0$
for $P(x,0)=\delta(x)$. It then follows that
\begin{equation}
\label{FMSD}
\frac{d^\alpha}{dt^\alpha}\langle x^2(t)\rangle=K_\alpha\int_{-\infty}^{\infty}
x^2\left[\frac{\partial^2}{\partial x^2}P(x,t)\right]dx=2K_\alpha.
\end{equation}
Applying a Laplace transformation to Eq.~(\ref{FMSD}),
\begin{equation}
\mathscr{L}\left[\frac{d^\alpha}{dt^\alpha}\langle x^2(t)\rangle\right]=2K_
\alpha\frac{1}{s}.
\end{equation}

For the Caputo-derivative,
\begin{equation}
s^\alpha\left\langle\tilde{x}^2(s)\right\rangle_{C} =2 D_\alpha \frac{1}{s},
\end{equation}
and the second moment becomes
\begin{equation}
\langle x^2(t)\rangle=\frac{2K_\alpha}{\Gamma(\alpha+1)}t^\alpha,
\end{equation}
which is a familiar results \cite{report}.

For the CF-GDE,
\begin{equation}
\frac{s}{(1-\alpha)s+\alpha}\langle\tilde{x}^2(s)\rangle_{CF}=2K_\alpha
\frac{1}{s},
\end{equation}
such that
\begin{equation}
\langle\tilde{x}_{(s)}^2\rangle=2K_\alpha(1-\alpha)\frac{1}{s}+2K_\alpha
\alpha\frac{1}{s^2},
\end{equation}
and then
\begin{equation}
\langle x^2(t)\rangle_{CF}=2K_\alpha(1-\alpha)+2K_\alpha\alpha t.
\end{equation}

For the AB-GDE,
\begin{equation}
\frac{s^\alpha}{(1-\alpha)s^\alpha+\alpha}\langle\tilde{x}^2(s)\rangle=2
K_\alpha\frac{1}{s},
\end{equation}
such that
\begin{equation}
\langle\tilde{x}^2(s)\rangle=2K_\alpha(1-\alpha)\frac{1}{s}+2K_\alpha\alpha
\frac{1}{s^{\alpha+1}},
\end{equation}
and then
\begin{equation}
\langle x^2(t)\rangle_{AB}=2K_\alpha(1-\alpha)+2K_\alpha\frac{t^\alpha}{
\Gamma{(\alpha)}}.
\end{equation}

For the kurtosis, we calculate the third and fourth order moments,
\begin{eqnarray}
\langle x^3(s)\rangle&=&-i\left.\frac{\partial^3}{\partial k^3}\hat{\tilde{
P}}(k,s)\right|_{k=0},\\
\langle x^4(s)\rangle&=&\left.\frac{\partial^4}{\partial k^4}\hat{\tilde{P}}(
k,s)\right|_{k=0}.
\end{eqnarray}
From these we find the kurtosis
\begin{equation}
\kappa(t)=\left<\left(\frac{x-\langle x\rangle}{\left<\left(x-\langle x\rangle
\right)^2\right>^{1/2}}\right)^4\right>.
\end{equation}
The kurtosis for the CF- and AB-GDEs are
\begin{equation}
\label{KCF}
\kappa_{CF}=6\left[\frac{(1-\alpha)+\alpha\frac{t}{\sqrt2}}{(1-\alpha)+\alpha
t}\right]^2,
\end{equation}
and
\begin{equation}
\label{KAB}
\kappa_{AB}=6\frac{\frac{\alpha t^{2\alpha}}{2\Gamma{(2\alpha)}}+2(1-\alpha)
\frac{t^\alpha}{\Gamma{(\alpha)}}+(1-\alpha)^2}{\left[\frac{t^\alpha}{
\Gamma{(\alpha)}}+(1-\alpha)\right]^2}.
\end{equation}

\subsection{Integro-differential operators in the Langevin equation}
\label{AppendixB2}

The generalized Langevin equation with integro-differential operators is
\begin{equation}
\frac{d^{\alpha}}{dt^{\alpha}}x(t)=\sqrt{2K_{\alpha}}\xi(t),
\end{equation}
where $0<\alpha\leq1$ and $d^{\alpha}/dt^{\alpha}$ represents the Caputo-,
CF- and AB-operators. Applying a Laplace transformation,
\begin{equation}
\label{eq111}
\tilde{x}(s)=\sqrt{2K_{\alpha}}\frac{1}{s\tilde{\theta}(s)}\tilde{\xi}(s),
\end{equation}
where $\theta_{C}(t)=\frac{t^{-\alpha}}{\Gamma(1-\alpha)}$, $\theta_{CF}(t)=\frac{1}{(1-\alpha)\tau^{\alpha}}\exp{\left(-\frac{\alpha t}{(1-\alpha)\tau}\right)}$, 
$\theta_{AB}(t)=\frac{1}{(1-\alpha)\tau^{\alpha}}E_{\alpha}\left(-\alpha\frac{t^{\alpha}}{(1-\alpha)\tau^{\alpha}}\right)$.
After inverse Laplace transformation, we obtain
\begin{equation}
x(t)=\sqrt{2K_{\alpha}}\int_0^tH(t-t')\xi(t')dt'.
\end{equation}
The MSD is then
\begin{equation}
\langle x^2(t)\rangle=2K_{\alpha}\int_0^tH^2(t')dt'.
\end{equation}
Here, $H_{CF}(t)=(1-\alpha)\tau^{\alpha}\delta(t)+\alpha\tau^{\alpha-1}$, and
$H_{AB}(t)=(1-\alpha)\tau^{\alpha}\delta(t)+\alpha\frac{t^{\alpha-1}}{\Gamma(\alpha)}$.

For the Caputo derivative,
\begin{equation}
x(t)=\sqrt{2K_{\alpha}}\int_0^t\frac{(t-t')^{\alpha-1}}{\Gamma(\alpha)}\xi(t')
dt'.
\end{equation}
The two-point correlation function for the Caputo-fractional Langevin equation
is
\begin{eqnarray}
\nonumber
\langle x(t_1)x(t_2)\rangle&=&2K_{\alpha}\int_0^{t_1}\frac{(t_1-t_1')^{\alpha
-1}}{\Gamma(\alpha)}dt_1'\\
\nonumber
&&\hspace*{-1.2cm}\times\int_0^{t_2}\frac{(t_2-t_2')^{\alpha-1}}{\Gamma
(\alpha)}dt_2'\langle\xi(t_1')\xi(t_2')\rangle\\
\nonumber
&&\hspace*{-1.8cm}=\frac{2K_{\alpha}}{\Gamma{(\alpha)}^2}\int_0^{t_1}(t_1-t_1')
^{\alpha-1}dt_1'\\
\nonumber
&&\hspace*{-1.2cm}\times\int_0^{t_2}(t_2-t_2')^{\alpha-1}dt_2'\delta(t_1'-t_2')\\
\nonumber
&&\hspace*{-1.8cm}=\frac{2K_{\alpha}}{\Gamma{(\alpha)}^2}\int_0^{t_2}(t_1-t_2')^{\alpha-1}
(t_2-t_2')^{\alpha-1}dt_2'\\
\nonumber
&&\hspace*{-1.8cm}=\frac{2K_{\alpha}}{\Gamma{(\alpha)}^2}t_2^{\alpha}t_1^{\alpha-1}\int_0^1
(1-\frac{t_2}{t_1}x)^{\alpha-1}(1-x)^{\alpha-1}dx\\
&&\hspace*{-1.8cm}=\frac{2K_{\alpha}t_2^{\alpha}t_1^{\alpha-1}}{\alpha\Gamma{
(\alpha)}^2}\,_2F_1\left(1-\alpha,1;\alpha+1;\frac{t_{2}}{t_{1}}\right).
\label{A14}
\end{eqnarray}
Without restricting generality, we here assume that $t_1\ge t_2$, and
\begin{equation}
_2F_1(a,b;c;z)=\frac{\Gamma(c)}{\Gamma(b)\Gamma(c-b)}\int_0^1\frac{t^{b-1}
(1-t)^{c-b-1}}{(1-tz)^a}dt
\end{equation}
is the hypergeometric function, which for $|z|<1$ is defined by the power
series
\begin{equation}
\label{hyper}
_2F_1(a,b;c;z)=\sum_{n=0}^{\infty}\frac{(a)_n(b)_n}{(c)_n}\frac{z^n}{n !}.
\end{equation}
Here $(q)_n$ is the (rising) Pochhammer symbol
\begin{equation*}
(q)_n=\left\{\begin{array}{ll}1,&n=0\\q(q+1)\cdots(q+n-1),&n>0\end{array}
\right..
\end{equation*}

Then the MSD is
\begin{equation}
\langle x^2(t)\rangle=C_1t^{2\alpha-1},
\end{equation}
for $\alpha>1/2$, and where $C_1=2K_\alpha/[(2\alpha-1)\Gamma^2(\alpha)]$.

\section{TAMSD and ACVF for Caputo Langevin equation}
\label{AppendixC}

\subsection{TAMSD}
\label{AppendixC1}

According to the definition (\ref{TAMSD}) of the TAMSD, the mean TAMSD of the
Caputo-fractional Langevin equation can be derived in the form
\begin{eqnarray}
\nonumber
\left<\overline{\delta^2(\Delta)}\right>&=&\left<\frac{1}{T-\Delta}\int_0^{T-
\Delta}\left[x(t+\Delta)-x(t)\right]^2dt\right>\\
\nonumber
&&\hspace*{-1.8cm}=\frac{1}{T-\Delta}\int_0^{T-\Delta}\left[\langle x^2(t+
\Delta)\rangle+\langle x^2(t)\rangle-2I_1\right]dt\\
&&\hspace*{-1.8cm}=\frac{C_1}{2\alpha}\frac{1}{T-\Delta}\left[T^{2\alpha}-
\Delta^{2\alpha}+(T-\Delta)^{2\alpha}\right]-2I,
\label{CTAMSD}
\end{eqnarray}
where $C_1=2K_\alpha/[(2\alpha-1)\Gamma^2(\alpha)]$ and $I=(T-\Delta)^{-1}
\int_0^{T-\Delta}I_1dt$. In this latter expression we used
\begin{eqnarray}
\nonumber
I_1&=&\langle x(t+\Delta)x(t)\rangle=C_2(t+\Delta)^{\alpha-1}t^\alpha\\
&&\times\,_2F_1\left(1-\alpha,1;1+\alpha;\frac{t}{t+\Delta}\right),
\label{I11}
\end{eqnarray}
with $C_2=2K_\alpha/[\alpha\Gamma^2(\alpha)]$. We note that in particular
when $\Delta=0$, we get $\langle\overline{\delta^2(0)}\rangle=0$.

Now we focus on the case when $\Delta\neq0$, and, more specifically, on the
limit $\Delta/T\ll1$. We first calculate $I$ in Eq.~(\ref{CTAMSD}), using
Eq.~(15.3.4) in  Ref.~\cite{abramowitz1964handbook} for $I_{1}$ in
Eq.~(\ref{I11}). Then
\begin{equation}
\label{I1}
I_1=C_2\Delta^{\alpha-1}t^\alpha\,_2F_1\left(1-\alpha,\alpha;\alpha+1;
-\frac{t}{\Delta}\right).
\end{equation}
Using the relation between the $H$-function and the hypergeometric functions,
Eq.~(1.131) in  Ref.~\cite{mathai}, we find
\begin{eqnarray}
\nonumber
_2F_1\left(1-\alpha,\alpha;\alpha+1;-\frac{t}{\Delta}\right)&=&\frac{\alpha}{
\Gamma(1-\alpha)}\\
&&\hspace*{-3.2cm}\times H_{2,2}^{1,2}\left[\frac{t}{\Delta}\left|\begin{array}{l}
(\alpha,1),(1-\alpha,1)\\(0,1),(-\alpha,1)\end{array}\right.\right].
\end{eqnarray}
Applying relation 1.16.4 in  Ref.~\cite{prudnikov1990more}, we obtain
\begin{eqnarray}
\nonumber
I&=&\frac{C\Delta^{2\alpha}}{(T-\Delta)}\int_0^{T^*}s^\alpha H_{2,2}^{1,2}
\left[s\left|\begin{array}{l}(\alpha,1),(1-\alpha,1)\\(0,1),(-\alpha,1)
\end{array}\right.\right]ds\\
\nonumber
&=&\frac{C(T^*)^{\alpha}}{\Delta^{1-2\alpha}}\\
&&\times H_{3,3}^{1,3}\left[T^*\left|
\begin{array}{l}(-\alpha,1),(\alpha,1),(1-\alpha,0)\\(0,1),(-\alpha,1),
(-\alpha-1,1)\end{array}\right.\right],
\end{eqnarray}
where $T^*=(T-\Delta)/\Delta$ and $C=\alpha C_2/\Gamma(1-\alpha)$. Using
relation 8.3.2.7 in Ref.~\cite{prudnikov1990more}, we then obtain
\begin{equation}
I=\frac{C(T^*)^{\alpha}}{\Delta^{1-2\alpha}}H_{3,3}^{3,1}\left[\frac{1}{T^*}
\left|\begin{array}{l}(1,1),(1+\alpha,1),(2+\alpha,1)\\(1+\alpha,1),(1-\alpha,
1),(\alpha,1)\end{array}\right.\right].
\end{equation}
With relation 8.3.2.3 in Ref.~\cite{prudnikov1990more}, in the limit
$\Delta/T\ll1$ we get
\begin{eqnarray}
\nonumber
I&\sim&\frac{K_{\alpha}}{\alpha(2\alpha-1)\Gamma^2(\alpha)}(T-\Delta)^{2\alpha-1}\\
\nonumber
&&+\frac{K_{\alpha}}{\Gamma(2\alpha)\cos(\pi\alpha)}\Delta^{2\alpha-1}\\
&&+\frac{K_{\alpha}}{(2\alpha-1)\Gamma^2(\alpha)}\Delta(T-\Delta)^{2\alpha-2}.
\end{eqnarray}
We then obtain the leading behavior of the mean TAMSD
of the Caputo-fractional Langevin equation (\ref{CTAMSD}),
\begin{equation}
\left<\overline{\delta^2(\Delta)}\right>\sim\frac{2K_{\alpha}}{\Gamma(2\alpha)
\left|\cos(\pi\alpha)\right|}\Delta^{2\alpha-1}.
\end{equation}

\subsection{ACVF}
\label{AppendixC2}

From the two-point correlation function (\ref{CaputoLEtwo}) for the
Caputo-Langevin equation, we here calculate the ACVF. The general result
with $t>0$ is
\begin{eqnarray}
\nonumber
&&C_{\Delta}(t)=\frac{2K_{\alpha}}{\alpha\Gamma^2(\alpha)\Delta^2}\\
\nonumber
&&\hspace*{0.8cm}\times\left[\Delta^{\alpha}(t+\Delta)^{\alpha-1}{}_2F_1
\left(1-\alpha,1,\alpha+1,\frac{\Delta}{t+\Delta}\right)\right.\\
&&\hspace*{1.4cm}\left.-\Delta^{\alpha}t^{\alpha-1}{}_2F_1\left(1-\alpha,1,
\alpha+1,\frac{\Delta}{t}\right)\right]. 
\end{eqnarray}
Using Eq.~(15.3.4) in Ref.~\cite{abramowitz1964handbook} to the hypergeometric
function $_2F_1(1-\alpha,1,\alpha+1,\Delta/[t+\Delta])$, we obtain
\begin{eqnarray}
\nonumber
&&C_{\Delta}(t)=\frac{2K_{\alpha}}{\alpha\Gamma^2(\alpha)\Delta^2}t^{\alpha-
1}\Delta^{\alpha}\\
\nonumber
&&\hspace*{0.8cm}\times\left[{}_2F_1\left(1-\alpha,\alpha,\alpha+1,-\frac{
\Delta}{t}\right)\right.\\
&&\hspace*{1.4cm}\left.-{}_2F_1\left(1-\alpha,1,\alpha+1,\frac{\Delta}{t}
\right)\right].
\end{eqnarray}
When $t=0$, $C_{\Delta}(0)=\langle x^2(\Delta)\rangle/\Delta^2=\frac{2
K_{\alpha}}{(2\alpha-1)\Gamma^2(\alpha)}\Delta^{2\alpha-3}$.
Moreover, when $t\gg\Delta$, using (\ref{hyper}) we have
\begin{equation}
_2F_1\left(1-\alpha,\alpha,\alpha+1,-\frac{\Delta}{t}\right)\sim1-\frac{\alpha(
1-\alpha)}{\alpha+1}\frac{\Delta}{t},
\end{equation}
and
\begin{equation}
_2F_1\left(1-\alpha,1,\alpha+1,\frac{\Delta}{t}\right)\sim1+\frac{1-\alpha}{
\alpha+1}\frac{\Delta}{t},
\end{equation}
and then
\begin{equation}
C_{\Delta}(t)\sim\frac{2(\alpha-1)K_{\alpha}}{\alpha\Gamma^2(\alpha)}\Delta
^{\alpha-1}t^{\alpha-2}.
\end{equation}

\section{Simulations}
\label{AppendixA}

We summarize the discretization scheme for the Langevin equation with
Caputo-fractional and conformable derivatives, as well as for SBM.

\subsection{Caputo Langevin Equation}

We apply the implicit difference method \cite{gu2020meshless}. Let $t=\left[t_0,
t_{n+1}\right]$ with uniform step size $\Delta t=t_{k+1}-t_k$, $k=1,2,\ldots,n$.
Then the left side of Eq.~(\ref{caputoLE}) with Caputo derivative is reduced to
\begin{equation}
\frac{d^\alpha x_{n+1}}{dt^\alpha}=\frac{1}{\Gamma(1-\alpha)}\int_{t_0}^{t_{n+1}}
\frac{dx(\mu)}{d\mu}\left(t_{n+1}-\mu\right)^{-\alpha}d\mu,
\end{equation}
or
\begin{equation}
\label{52}
\frac{d^\alpha x_{n+1}}{dt^\alpha}=\frac{1}{\Gamma(1-\alpha)}\sum_{i=0}^n
\int_{t_i}^{t_{i+1}}\frac{dx(\mu)}{d\mu}\left(t_{n+1}-\mu\right)^{-\alpha}d\mu,
\end{equation}
where
$dx(\mu)/d\mu$ can be approximated by the implicit difference method as
\begin{equation}
\label{53}
\frac{dx(\mu)}{d\mu}=\frac{x_{i+1}-x_i}{\Delta t}+O(\Delta t),
\end{equation}
where $\mu\in[t_i,t_{i+1}]$. The remaining integral terms can be solved via
\begin{eqnarray}
\nonumber
\int_{t_i}^{t_{i+1}}(t_{n+1}-\mu)^{-\alpha}d\mu&=&\frac{(\Delta t)^{1-\alpha}}{
1-\alpha}\\
&&\hspace*{-2cm}\times[(n-i+1)^{1-\alpha}-(n-i)^{1-\alpha}].
\label{54}
\end{eqnarray}
Let $a_{n-i}=(n-i+1)^{1-\alpha}-(n-i)^{1-\alpha}$ and $a_0=1$. Substituting
Eqs.~(\ref{53}) and (\ref{54}) into (\ref{52}), one then has
\begin{eqnarray}
\nonumber
\frac{d^\alpha x_{n+1}}{d t^\alpha}&\simeq&\frac{(\Delta t)^{-\alpha}}{\Gamma(
2-\alpha)}\sum_{i=0}^n\left(x_{i+1}-x_{i}\right)\\
\nonumber
&&\times\left[(n-i+1)^{1-\alpha}-(n-i)^{1-\alpha}\right]\\
\nonumber
&=&\frac{(\Delta t)^{-\alpha}}{\Gamma(2-\alpha)}\left[-\sum_{i=1}^n x_i(a_{n-i}
-a_{n-i-1})\right.\\
&&\left.-x_0 a_n+a_0x_{n+1}\right].
\end{eqnarray}

The right hand side of Eq.~(\ref{caputoLE}) with Caputo-fractional derivative
is $\sqrt{2K_{\alpha}/\Delta t}\eta(i)$, where $\eta(i)$ is a zero-mean
Gaussian random variable with unit standard deviation. Then
\begin{eqnarray}
\nonumber
&&-\frac{(\Delta t)^{-\alpha}}{\Gamma(2-\alpha)}\left[\sum_{i=1}^nx_i(a_{n-i}
-a_{n-i-1})+x_0 a_n+a_0 x_{n+1}\right]\\
&&\hspace*{1.8cm}=\sqrt{\frac{2D_{\alpha}}{\Delta t}}\eta(i),
\end{eqnarray}
and
\begin{eqnarray}
\nonumber
x_{n+1}&=&\sqrt{2K_\alpha}(\Delta t)^{\alpha-\frac{1}{2}}\Gamma(2-\alpha)
\eta(i)\\
&&+x_0 a_n+\sum_{i=1}^n x_i(a_{n-i}-a_{n-i-1}).
\end{eqnarray}
For $x_j$, $j=1,2,\ldots,n$,
\begin{eqnarray}
\nonumber
x_j&=&\sqrt{2K_\alpha}(\Delta t)^{\alpha-\frac{1}{2}}\Gamma(2-\alpha)
\eta(i)\\
&&+x_0a_{j-1}+\sum_{i=1}^{j-1}x_i(a_{j-i-1}-a_{j-i-2}).
\end{eqnarray}

\subsection{SBM- and conformable-Langevin equation}

The Langevin equation of SBM is
\begin{equation}
\frac{dx(t)}{dt}=\sqrt{2K_\alpha(t)}\xi(t).
\end{equation}
With Eq.~(\ref{53}) we deduce that
\begin{equation}
\frac{x_{i+1}-x_i}{\Delta t}=\sqrt{\frac{2\alpha K_\alpha i^{\alpha-1}}{\Delta
t}}(\Delta t)^{\frac{\alpha-1}{2}}\eta(i),
\end{equation}
that is,
\begin{equation}
x_{i+1}=x_i+\sqrt{2\alpha K_\alpha i^{\alpha-1}}(\Delta t)^{\frac{\alpha}{2}}
\eta(i)
\end{equation}

For the conformable-Langevin equation (\ref{eq20}), the finite difference
method produces
\begin{equation}
\alpha\frac{x(t_1)-x(t_2)}{t_1^\alpha-t_2^\alpha}=\sqrt{2K_\alpha}\xi(t_1),
\end{equation}
and we finally have
\begin{equation}
x_{i+1}=x_i+\frac{\sqrt{2K_{\alpha}}}{\alpha}(\Delta t)^{\alpha-\frac{1}{2}}
\left[(i+1)^\alpha-i^\alpha\right]\eta(i) .
\end{equation}

\end{document}